\documentclass[showpacs,preprintnumbers,amsmath,amssymb,eqsecnum]{revtex4}

\usepackage{graphicx}
\usepackage{dcolumn}
\usepackage{bm}
\usepackage{subfigure}

\begin{document}


\title{Discretisation parameter and operator ordering in 
loop quantum cosmology with  the cosmological constant
}

\author{$^{1}$Tomo Tanaka}
\email{tomo@gravity.phys.waseda.ac.jp}

\author{$^{2}$Fumitoshi Amemiya}
\email{famemiya@rk.phys.keio.ac.jp}

\author{$^{3}$Masahiro Shimano}%
 \email{shimano@rikkyo.ac.jp}
\author{$^{3}$Tomohiro Harada }%
 \email{harada@rikkyo.ac.jp}
\author{$^{4}$Takashi Tamaki}
\email{tamaki@ge.ce.nihon-u.ac.jp}
\affiliation{%
$^{1}$Department of Physics, Waseda University, Okubo, Tokyo 169-8555, Japan\\
$^{2}$Department of Physics, Keio University, Hiyoshi, Kohoku-ku, Yokohama 223-8522, Japan\\
$^{3}$Department of Physics, Rikkyo University, Toshima, Tokyo 175-8501, Japan
\\
$^{4}$College of Engineering, Nihon University, Koriyama 963-8642, Japan
}%

\date{\today}

\begin{abstract}
In loop quantum cosmology, the Hamiltonian reduces to a finite difference operator. 
We study the initial singularity and the large volume limit
against the ambiguities in the discretisation and the operator ordering within 
a homogeneous, isotropic and spatially flat model with the cosmological constant. 
We find that the absence of the singularity strongly depends on the choice 
of the operator ordering and the requirement for the absence singles out a very small class
of orderings. 
Moreover we find a general ordering rule required for the absence of the singularity. 
We also find that the large volume limit naturally recovers a smooth 
wave function in the discretisation where each step corresponds to a fixed volume 
increment but not in the one where each step corresponds to a fixed area increment. 
If loop quantum cosmology is to be a phenomenological realisation of full loop quantum gravity, 
these results are important to fix the theoretical ambiguities. 
\end{abstract}

\maketitle

\section{Introduction}
Loop quantum gravity (LQG) is a background-independent nonperturbative quantum gravity. 
There, one quantises the Hamiltonian formulation of general relativity 
based on the SU(2) connection $A^{i}_{a}$ and the densitised triad $E^{a}_{i}$
on the three-dimensional space, where $a$, $b$, ... and $i$, $j$, ... both run over 
1, 2 and 3. $a$, $b$, ... are tangent indices, while $i$, $j$, ... 
are indices associated with a basis $\{\tau_{i}\}$ for the 
Lie algebra of SU(2) Lie group and raised and lowered 
with Kronecker's delta $\delta_{ij}$. The three-dimensional inverse metric is given by 
$q^{ab}=|\det E^{c}_{k}|^{-1}E^{a}_{i}E^{b}_{j}\delta^{ij}$.
To quantise the system, LQG invokes holonomies $h_{e}=e^{\int_{e}A^{i}_{a}\tau_{i}dx^{a}}$
along an edge $e$ and fluxes ${\cal E}_{\cal S}=\int_{\cal S} d^{2}\sigma n_{a} E^{a}_{i}\tau^{i}$
over a two-surface ${\cal S}$, where $d^{2}\sigma$ is the area element on ${\cal S}$ and $n_{a}$ is a unit normal
to ${\cal S}$. There appears a nondimensional constant 
parameter $\gamma$, which is called the Barbero-Immirzi parameter
 and cannot be determined within the theory. 
The kinematical Hilbert space is spanned by the spin network states.
The Hamiltonian as well as the area and the volume are constructed 
from holonomies and fluxes and act as operators on the kinematical Hilbert space. 
The Hilbert space is spanned by equivalence classes of 
the spin networks under diffeomorphisms or the s-knot states. 
 See~\cite{Rovelli0, Thiemann0, Ashtekar_2005} for 
the basics and recent developments in LQG.

Loop quantum cosmology (LQC)~\cite{lqc} is a quantised 
minisuperspace model motivated by LQG. 
In traditional quantum cosmology~\cite{traditional_QC}, 
symmetry reduced models are quantised in the usual Schr\"{o}dinger representation
and the quantised Hamiltonian constraint yields the Wheeler-DeWitt (WDW) differential equation,
while in LQC one uses the so-called polymer particle representation~\cite{polymer}
which is unitary inequivalent to the Schr\"{o}dinger representation
and obtains a second order {\it difference equation} rather than the differential equation.
Remarkably, Bojowald~\cite{Bojowald4} demonstrated that 
there can be no big bang singularity in the following two aspects. 
First, 
the spectrum of the inverse scale factor operator is bounded from above.
Second, the wave function of the universe can be uniquely extended 
beyond the point which was the initial singularity in classical theory.
The above features have been first shown for a homogeneous, isotropic and 
spatially flat model and subsequently generalised to nonflat or 
anisotropic cases~\cite{lqc}, 
although it is yet uncertain
whether this can be generalised to inhomogeneous models
(see e.g.~\cite{bojowald_etal_2008,bojowald_etal_2009}).

In this paper, we visit the absence of the singularity 
in the presence of the cosmological constant against the 
quantisation ambiguities. In particular, we focus on the choice of 
the discretisation parameter $\lambda$ which corresponds to the coordinate 
length of edges for the basic holonomies and 
the operator ordering in the Hamiltonian 
constraint. To study these issues, we fix the cosmological model to be 
homogeneous, isotropic and spatially flat with the cosmological constant. 
We first adopt the constant $\lambda$ as in~\cite{Bojowald4}, for which the universe 
gains a fixed area quantum at each step and find that 
the absence of the singularity strongly depends on the 
operator ordering in the Hamiltonian. 
However, for this discretisation, we encounter a serious problem 
in the physical interpretation of the 
obtained wave function in the large volume limit. 
The problem is that the behaviour of the wave function obtained from the difference equation 
in LQC does not agree with that obtained from the WDW equation in the large volume limit.
Since the effects of discreteness are dominant in the Planck scale physics but 
should disappear when the universe becomes large,
the discrete wave function in LQC in the large volume is expected to have a correspondence to 
the smooth wave function which is a solution to the WDW equation.
We see that this problem is resolved if $\lambda$
is chosen to vary~\cite{Ashtekar} so that the universe gains a fixed 
volume quantum and find that the absence of the singularity strongly depends on the 
operator ordering in the Hamiltonian also in this discretisation.
It is already studied in detail how the large volume limit depends on the choice of the 
discretisation in the presence of matter fields by Nelson and 
Sakellariadou~\cite{Nelson_Sakellariadou_2007a,Nelson_Sakellariadou_2007b}
and the present result with the cosmological constant but without matter fields
is consistent with theirs.

This paper is organised as follows. In Section~\ref{sec:lqc}, we introduce 
the discretised Hamiltonian for a homogeneous, isotropic and spatially flat model with the 
cosmological constant. 
In Section~\ref{sec:equi-area}, we choose $\lambda$ to be constant and derive the 
difference equation. Then, we demonstrate how the absence of the singularity
depends on the choice of the operator ordering and numerically and analytically
show the large volume limit is problematic. 
In Section~\ref{sec:equi-volume}, we vary $\lambda$ appropriately and derive the 
difference equation. Then, we
show that the large volume limit problem is resolved in this discretisation
and also find that the absence of the singularity strongly 
depends on the choice of the operator ordering. 
Section~\ref{sec:discussion} 
is devoted to discussion for the influence of the matter fields.
In Section~\ref{sec:conclusion} we conclude the paper.
In this paper we use the units in which $c=\hbar=1$.

\section{Loop quantum cosmology}
\label{sec:lqc}
\subsection{Hamiltonian constraint}

In the present paper, we focus on a homogeneous, isotropic and spatially
flat universe. In classical theory, the line element for such a spacetime
is given by the flat Friedmann-Robertson-Walker (FRW) metric
	\begin{equation}
	 ds^2=-dt^2+a(t)^2\left(dx^2+dy^2+dz^2\right),
	\end{equation}
where $a(t)$ is called the scale factor.
To remove the divergence coming from the volume integral we introduce an elementary cell 
{\Large $\nu$} on the three dimensional space. 
In this case, the gravitational Hamiltonian constraint is written as~\cite{Bentivegna}
	\begin{equation}\label{hamiltonian}
         C_{\mathrm{grav}}=\frac{1}{16\pi G\gamma^2}\int_{\mathcal{\nu}}d^3x
         \left(-\frac{1}{\sqrt{|\det E^a_i|}}\epsilon_{ijk}F^i_{ab}E^{aj}E^{bk}+2\gamma^2\sqrt{|\det E^a_i|}\Lambda\right), 
        \end{equation}
where $G$, $\Lambda$, $\epsilon_{ijk}$ and 
$F^i_{ab}=\partial_aA^i_b-\partial_bA^i_a+\epsilon^i_{jk}A^j_aA^k_b$
are the gravitational constant, 
the cosmological constant, 
the Levi-Civita symbol
and the curvature associated with the connection $A^{i}_{a}$, 
respectively. Although the cosmological constant 
might be an emergent object from some unknown effects,
we incorporate it into the Hamiltonian as usual for simplicity.
In the elementary cell {\Large $\nu$}, we define a time-independent 
fiducial flat metric $ \bar{q}_{ab}$, an associated orthonormal triad $\{ \bar{e}^a_i \}$ and cotriad $\{ \bar{\omega}_a^i \}$.
In the flat FRW universe, the connection $A_a^i$ and the densitised triad 
$E^a_i$ are given by 
	\begin{align} 
	 A^i_a =cV_0^{-(1/3)} {}\bar{\omega}^i_a, \quad 
         E^a_i = p V_0^{-(2/3)} \sqrt{ \bar{q}} {}\bar{e}^a_i,
	\label{eq:canonical_variables}
\end{align} 
where $\bar{q}$ is the determinant of $\bar{q}_{ab}$, $c= V_0^{1/3}{\rm sgn}(p)\gamma da/dt$, 
$\vert p\vert =V_0^{2/3}a^2$ and $V_0=\int_{\large \nu}d^{3}x \sqrt{\bar{q}}$. $V_{0}$ is thus the volume of the elementary cell with respect to the fiducial metric.
In other words, $|p|$ is proportional to the physical area of the elementary cell 
and $c$ is the conjugate momentum of $p$. 
The Poisson bracket between $c$ and $p$ takes the form $\{ c,p \} = 8\pi G\gamma/3$.
Now, the holonomy $h_i^{(\lambda)}$ along an edge parallel to
$\bar{e}^a_i$ is given by
	\begin{align}
	 h_i^{(\lambda)} = e^{\lambda c\tau_{i}}, 
	\label{holonomy}
	\end{align}
where $\lambda$ is the coordinate length of the edge and $\{\tau_i\}$ are 
a basis for the Lie algebra of SU(2) Lie group. 
The flux ${\cal E_{\cal S}}$ is simply given by ${\cal E_{\cal S}}=p V_{0}^{-2/3}A_{\cal S}$, 
where $A_{\cal S}$ is the area of the surface ${\cal S}$~\cite{Ashtekar0}.
Note that 
in defining the Hamiltonian constraint operator, 
one traces over SU(2) valued holonomies, 
and then there appears an ambiguity in choosing the irreducible representation to perform the trace.
Here, we choose the spin $J=1/2$ representation for simplicity. 
Then the holonomy becomes
	\begin{equation}
         h_i^{(\lambda)}=\cos\left(\lambda c/2\right) \mathbb{I} + 2\sin\left(\lambda c/2\right) \tau_i,
        \end{equation}
where $\mathbb{I}$ is the unit $2\times2$ matrix 
and $\{\tau_i\}$ are related to the Pauli matrices $\{\sigma_i\}$ via $2i\tau_i=\sigma_i$.

We rewrite the Hamiltonian constraint in terms of the flux $p$ and the holonomy $h_k^{(\lambda)}$ 
as in the full theory. 
For the triad part of the first term in the constraint (\ref{hamiltonian}), we obtain
	\begin{equation}\label{traiad_term}
	 \epsilon_{ijk}\tau^i\frac{E^{aj}E^{bk}}{\sqrt{|\det E^a_i|}}=
         -\frac{2\mathrm{sgn}(p)}{8\pi G\gamma \lambda V_0^{\frac{1}{3}}} \bar{\epsilon}^{abc}
         \bar{\omega}_c^i\left(h_i^{(\lambda)}
	 \left\{\left(h_i^{(\lambda)}\right)^{-1},V\right\}\right),
	\end{equation}
where $\bar{\epsilon}^{abc}$ is the Levi-Civita symbol, $V$ is the volume and 
$\{\bullet ,\bullet \}$ denotes the Poisson bracket. 
For the curvature $F^i_{ab}$, we rely on a standard prescription 
in gauge theories. We consider a loop $\alpha$, which is a square $\square_{ij}$ spanned by two triad vectors $\bar{e}^{a}_{i}$ and $\bar{e}^{b}_{j}$ of which each side is as long as $\lambda$ in the coordinate length. 
Then, the $ab$ component of the curvature is given by 
	\begin{equation} \label{curvature}
	 \tau_iF^i_{ab}=\lim_{\mathrm{Area}\rightarrow 0}\left(\frac{h_\alpha^{(\lambda)}{}_{ij}-\delta_{ij}}{\lambda^2 V_0^{\frac{2}{3}}}\right) \bar{\omega}^i_a \bar{\omega}^j_b,
	\end{equation}
where the holonomy $h_{\alpha}^{(\lambda)}{}_{ij}$ along $\alpha=\square_{ij}$ is the product of holonomies along the four edges, 
	\begin{equation}
	 h_\alpha^{(\lambda)}{}_{ij}=h_i^{(\lambda)}h_j^{(\lambda)}\left(h_i^{(\lambda)}\right)^{-1}\left(h_j^{(\lambda)}\right)^{-1}. \label{holonomy_square}
	\end{equation}
Substituting Eqs.~(\ref{traiad_term}), (\ref{curvature}), and (\ref{holonomy_square}) into Eq.~(\ref{hamiltonian}), 
$C_{\mathrm{grav}}$ can be expressed as 
	\begin{equation}\label{quantum operator}
         C_{\mathrm{grav}}= \frac{1}{16\pi G\gamma^2}
         \left(-\frac{4\mathrm{sgn}(p)}
         {8\pi \lambda^3G\gamma}\sum_{ijk}\epsilon^{ijk}\mathrm{Tr}\left[h_i^{(\lambda)}h_j^{(\lambda)}
         \left(h_i^{(\lambda)}\right)^{-1}\left(h_j^{(\lambda)}\right)^{-1} h_k^{(\lambda)}
         \left\{\left(h_k^{(\lambda)}\right)^{-1},V\right\}\right]+2\gamma^2 \Lambda V\right),
        \end{equation}
where we have used the relation 
	\begin{equation}
	 \tau_{i}\tau_{j}=\frac{1}{2}\epsilon_{ijk}\tau^{k}-\frac{1}{4}\delta_{ij}
	 \mathbb{I}.
	\end{equation}

To quantise the Hamiltonian, we replace $h^{(\lambda)}_{i}$ and $p$ with the 
corresponding operators. 
The kinematical Hilbert space is defined by
$\mathcal{H}_{\mathrm{kin}}^{\mathrm{grav}}=L^2(\mathbb{R}_{\mathrm{Bohr}}, \mathrm{d}\mu_{\mathrm{Bohr}})$,
where $\mathbb{R}_{\mathrm{Bohr}}$ is the Bohr compactification of the real line and 
$\mathrm{d}\mu_{\mathrm{Bohr}}$ is the Haar measure on it \cite{Thiemann0,Ashtekar_2005,polymer,Ashtekar}. 
An orthonormal basis for the kinematical Hilbert space is given by a set of 
eigenstates  $\{|\mu\rangle\}$ of $\widehat{p}$, 
which satisfy the orthonormality relations $\langle \mu_1 | \mu_2 \rangle = \delta_{\mu_1\mu_2}$.
The action of the triad operator $\widehat{p}$ on the state $|\mu\rangle$ is given by
	\begin{align}
	 \widehat{p}|\mu\rangle = \frac{8\pi \gamma l^2_{\mathrm{Pl}}}{6} \mu |\mu\rangle, 
	 \label{eq:phatmu}
	\end{align}
where $l_{\mathrm{Pl}}:=\sqrt{G}$ is the Planck length. 
That is, the eigenvalues of $\widehat{p}$ are labelled by the dimensionless parameter $\mu$.
The states $\{|\mu\rangle\}$ are also eigenstates of the volume operator $\widehat{V}=\widehat{|p|^{3/2}}$: 
	\begin{equation}
	 \widehat{V}\vert \mu \rangle = \widehat{|p|^{3/2}}\vert \mu\rangle= V_{\mu}\vert\mu\rangle,
	 \label{eq:volume_eigenvalue}
	\end{equation}
where 
	\begin{equation}
	 V_{\mu}=\left(\frac{8\pi \gamma l^2_{\mathrm{Pl}}}{6} |\mu|\right)^{3/2}.
	\end{equation}

Using Eqs.~(\ref{holonomy}) and (\ref{quantum operator}) and 
replacing the Poisson bracket with a commutator, after some calculation, 
we find
	\begin{equation}\label{eq:HCO}
	 \widehat{C}_{\mathrm{grav}}=\frac{1}{16\pi l_{\mathrm{Pl}}^2\gamma^2}
	 \left(\frac{96i\left(\mathrm{sgn}(p)\right)}{8\pi\gamma l^2_{\mathrm{Pl}}}
	 \widehat{\frac{1}{\lambda^{3}}}
	 \widehat{{\sin}^2\frac{\lambda c}{2}{\cos}^2\frac{\lambda c}{2}}
	 \widehat{\left[{\sin}\frac{\lambda c}{2}{V}{\cos}\frac{\lambda c}{2}-{\cos}\frac{\lambda c}{2}V{\sin}\frac{\lambda c}{2}\right]}+2\gamma^2\Lambda \widehat{V}\right),
	\end{equation}
where the operator ordering is fixed for simplicity. The ambiguity in the ordering will be discussed later. Note that $\lambda$ itself is an operator in general. It should be noted that in LQC we do not take the limit $\lambda\to 0$. 
Later we will discuss the physical motivation for this setting.

\subsection{Discretisation ambiguity}\label{subsec:discretisation ambiguity}

In full LQG, the geometry is quantised through
the area and volume operators.
To define the area operator $\widehat{{\bf A}}({\cal S})$ for a two-surface ${\cal S}$,
we partition ${\cal S}$ into $N$ small two-surfaces $\{ {\cal S}_{n}\}$ such that $\cup_{n} {\cal S}_{n}={\cal S}$. 
For sufficiently large $N$, there is $\{{\cal S}_{n}\}$ such that no ${\cal S}_{n}$ will contain
more than one intersection with the graph $\Gamma$ of the spin network $|\mathrm{S}\rangle$.
The sum over $n$ reduces to a sum over the intersection points between
${\cal S}$ and $\Gamma$ and is independent of $N$ for sufficiently
large $N$. 
Then, the action of the area operator becomes
	\begin{equation}
	 \widehat{{\bf A}}({\cal S})|\mathrm{S}\rangle=
	 4\pi \gamma l_{\rm Pl}^{2}\sum_{i}\sqrt{2j^{u}_{i}(j^{u}_{i}+1)+2j^{d}_{i}(j^{d}_{i}+1)-j^{t}_{i}(j^{t}_{i}+1)}|\mathrm{S}\rangle,
	\end{equation}
where $\{i\}$ label the intersection points between the graph $\Gamma$ and the two-surface ${\cal S}$, 
the indices $u$, $d$ and $t$ stand for the edges upward, downward and tangential to the $\mathcal{S}$, respectively, 
and the positive half integers $j^{u}_{i}$, $j^{d}_{i}$ and $j^{t}_{i}$ are the spins of the links labelled by $i$ \cite{area spectrum}.
Thus, there appears the smallest area $\Delta =2\sqrt{3}\pi \gamma l_{\rm Pl}^{2}$. 
For the volume operator $\widehat{V}$ for a three-volume ${\cal R}$,
we take a similar strategy. We partition
the three-volume ${\cal R}$ into cubes $\{{\cal R}_{n}\}$
of the coordinate volume $\epsilon^{3}$
and for sufficiently small $\epsilon$ no cube will contain more than one node.
The volume operator has a nontrivial action
only on nodes and hence it will no longer depend on the
value of $\epsilon$. Then, it turns out that
the spin network state is an eigenstate of
the volume operator and the eigenvalue is given by the sum over the
nodes which are contained in the three-volume ${\cal R}$ and
at least quadrivalent.
Similarly, we define the quantised Hamiltonian, which has a nontrivial action
only on nodes. Thus, for sufficiently small $\epsilon$,  the action of the quantised 
Hamiltonian will not depend on $\epsilon$.

In LQC, however, we leave the parameter $\lambda$ of the discretisation
nonzero finite. In fact, we are forced to do so
in the present formulation of LQC because there is no operator
corresponding to $c$. On the other hand, the physical results
seem to
depend on the choice of $\lambda$.
From this point of view, there is no first
principle within the formulation of LQC about how to choose
the nonzero finite value for the discretisation parameter $\lambda$.
We should probably fix the discretisation parameter $\lambda$
(and other quantisation ambiguities) in LQC so that LQC can
reproduce the features that full LQG should have.

To see how $\lambda$ is kept nonzero, we see the following relation:
	\begin{equation}
         \widehat{|p|}h^{(\lambda)}_{i}=
	 \frac{8\pi}{6}\gamma l_{\mathrm{Pl}}^2|\lambda|h^{(\lambda)}_{i}. 
	 \label{eq:abs(p)}
     	\end{equation}
Note that $|p|$ is the physical area of the elementary cell. 
Since the curvature in the Hamiltonian invokes the holonomy $h_{\alpha}^{(\lambda)}{}_{ij}$ 
along the square $\square_{ij}$, it would be reasonable to assume that each edge of this 
square is quantised so that the holonomy along each edge intersects the smallest area $\Delta$ of full LQG. 
This argument motivates us to choose $\lambda$ to be a constant $\mu_{0}$ of order unity.
This choice is adopted in Refs.~\cite{Ashtekar0,Ashtekar2,Bojowald4}.
The value for $\mu_{0}$ to fulfill this requirement exactly is $3\sqrt{3}/2$. 
The basic operator which appears in the Hamiltonian is $\widehat{\exp\left(i\mu_0c/2\right)}$ 
and this acts on $|\mu\rangle$ as follows:
	\begin{equation}\label{eq:periodic_operator00}
	 \widehat{e^{ i\mu_{0}{c}/2}} |\mu\rangle = |\mu+\mu_{0} \rangle.
	\end{equation}
This means that the eigenvalues of $\widehat{\mu}$ 
for the states which appear in the Hamiltonian constraint 
are spaced at constant intervals. 
Note that  as seen in Eq.~(\ref{eq:periodic_operator00}), 
the operator $\widehat{\exp\left(i\mu_0c/2\right)}$ 
drags the state $|\mu\rangle$ by the parameter length $\mu_{0}$ along the vector $\mathrm{d}/\mathrm{d}\mu$. 
Then, we can rewrite Eq.~(\ref{eq:periodic_operator00}) as 
	\begin{equation}\label{eq:periodic_operator}
	 \widehat{e^{ i\mu_{0}{c}/2}} |\mu\rangle = e^{\mu_0\left(d/d\mu\right)}|\mu \rangle. 
	\end{equation}
We call this choice of the parameter $\lambda$ the equi-area discretisation.

However, this is not the only possible choice. 
Since the curvature in the Hamiltonian invokes the holonomy $h_{\alpha}^{(\lambda)}{}_{ij}$ 
along the square $\square_{ij}$, it would also be reasonable to assume that the area
of this square is set to be the smallest area $\Delta $ of full LQG.
This motivates us to choose $\lambda$ to be a function $\lambda=\bar{\mu}(p)$ such that
	\begin{equation}
	 \bar{\mu}^2|p|=\Delta. \label{barmu}
	\end{equation} 
This gives another choice of the parameter $\lambda$. In this case, 
from Eq.~(\ref{eq:phatmu}) we find that $\bar{\mu}$ depends on $\mu$ 
as follows:
	\begin{equation}
	 \bar{\mu}^{2}|\mu|=\frac{3\sqrt{3}}{2}.
	 \label{eq:mubar2mu}
	\end{equation}
Similarly to Eq.~(\ref{eq:periodic_operator}), 
we can rewrite the operator $\exp\left(i\bar{\mu}c/2\right)$ as 
	\begin{equation}\label{eq:periodic_operator2}
	 \widehat{e^{ i\mu_{0}{c}/2}} |\mu\rangle = e^{\bar{\mu}\left(d/d\mu\right)}|\mu \rangle. 
	\end{equation}
If we introduce a variable $v$ satisfying 
	\begin{equation}
	 dv= \frac{1}{\bar{\mu}}d \mu,
	 \label{eq:dvdmu}
	\end{equation}
we can rewrite the right hand side of Eq.~(\ref{eq:periodic_operator2}) as 
	\begin{equation}
	 e^{\bar{\mu}\frac{\mathrm{d}}{\mathrm{d} \mu}}\vert \mu \rangle=e^{\frac{\mathrm{d}}{\mathrm{d} v}}\vert \mu \rangle.
	\end{equation}
This means that if we adopt the $v$-representation of the wave function,
the action of the Hamiltonian becomes simple. 
The explicit integration of Eq.~(\ref{eq:dvdmu}) using Eq.~(\ref{eq:mubar2mu})
yields
	\begin{equation}
         v=\mathrm{sgn}(\mu)K |\mu|^{\frac{3}{2}},
	\label{eq:vmu}
	\end{equation}
where $K=2\sqrt{2}/(3\sqrt{3\sqrt{3}})$. 
We should note that $v$ is proportional to the volume $V$. 
This choice is adopted in Ref.~\cite{Ashtekar}. 
In this case, it is more convenient to use the eigenstates $|v\rangle$ of the 
volume operator $\widehat{V}$ to see the action of the Hamiltonian constraint. 
The eigenstate $|v\rangle$ satisfies
	\begin{equation}
	 \widehat{V}|v\rangle=V_v|v\rangle, 
	\end{equation}
where 
	\begin{equation}
	 V_v=\left(\frac{8\pi}{6} \gamma l_{\mathrm{Pl}}^{2}\right)^{3/2}
	 \frac{|v|}{K}. 
	\end{equation}
The basic operator which appears in the Hamiltonian is 
$\widehat{e^{i \bar{\mu}c/2}}$ and 
its action on $|v\rangle$ is given by 
	\begin{equation}
	 \widehat{e^{i \bar{\mu}c/2}}|v\rangle = |v+1\rangle. 
	\end{equation}
Consequently, 
the Hamiltonian constraint involves the volume eigenstates of which the eigenvalues 
are equally spaced. In other words, the spectrum of the volume operator 
is distributed at equidistant intervals in volume. Thus, we will call this discretisation the equi-volume discretisation. Note that
since $\lambda=\bar{\mu}(p)$ depends on $p$, $\lambda$ should be treated as
an operator in quantum theory. 

\section{Equi-area discretisation}
\label{sec:equi-area}

\subsection{Absence of singularity and operator ordering}\label{difference_eq_pre-imp}
First we concentrate on the equi-area discretisation.
The holonomy operator acts on $|\mu \rangle $ as
	\begin{equation}\label{holonomyu}
	 \widehat{h_k^{(\mu_{0} )}}|\mu\rangle = \frac{1}{2} (|\mu+\mu_{0} \rangle + |\mu-\mu_{0} \rangle)
	 \mathbb{I}+ \frac{1}{i} (|\mu+\mu_{0} \rangle - |\mu-\mu_{0} \rangle) \tau_k. 
	\end{equation}
Even within this discretisation,
there are many possible choices of the operator ordering. 
To make the notation simple, we put 
	\begin{equation}
	 \widehat{F}=\widehat{{\sin}^2\frac{\mu_{0}  c}{2}{\cos}^2\frac{\mu_{0}  c}{2}}, \quad 
	 \widehat{EE}=\widehat{{\sin}\frac{\mu_{0}  c}{2}{V}{\cos}\frac{\mu_{0}  c}{2}-
	 {\cos}\frac{\mu_{0}  c}{2}{V}{\sin}\frac{\mu_{0}  c}{2}},
	\end{equation}
where $F$ and $EE$ relate to the squared holonomy $h_\alpha{}_{ij}$~(\ref{holonomy_square}) 
and the Poisson bracket $h_i\left\{(h_i)^{-1}, V\right\}$ in Eq.~(\ref{traiad_term}), respectively. 
For simplicity, hereafter we fix the ordering for the contents of $\widehat{F}$ and $\widehat{EE}$, 
and then there are two choices of the operator ordering in the Hamiltonian constraint operator 
as $\widehat{F}\widehat{EE}$ or $\widehat{EE}\widehat{F}$.
We first consider the operator ordering adopted in Ref.~\cite{Bojowald4}:
	\begin{eqnarray}\label{Hamiltonian_operator_pre}
	 \widehat{C}_{\mathrm{grav}}=\frac{1}{16\pi l_{\mathrm{Pl}}^2\gamma^2}
	 \left(\frac{96i\left(\mathrm{sgn}(p)\right)}{8\pi\gamma l^2_{\mathrm{Pl}}\mu_{0} ^{3}}
	 \widehat{F}\widehat{EE}+2\gamma^2\Lambda\widehat{V}\right).
	\end{eqnarray}
Then the action of the operator on a state $|\Psi\rangle$ leads 
to the difference equation, i.e., $\langle \mu |\widehat{C}_{\mathrm{grav}}|\Psi\rangle=0$ yields
	\begin{align}\label{diff eq_Bojo+CC}
	 \left| V_{\mu+5\mu_{0} }-V_{\mu+3\mu_{0} }\right| \Psi(\mu+4\mu_{0} )  
         - \left(2\left| V_{\mu+\mu_{0} } - V_{\mu-\mu_{0} }\right| -\frac{16\pi\gamma^3 l^2_{\mathrm{Pl}}\mu_{0} ^{3}}{3}\Lambda V_{\mu}\right)\Psi(\mu)
          + \left| V_{\mu-3\mu_{0} }-V_{\mu-5\mu_{0} }\right| \Psi(\mu-4\mu_{0} )=0,
	\end{align}
where $\Psi(\mu)=\langle \mu|\Psi \rangle $.
If we interpret the triad coefficient $p$ as an internal time, 
we can regard the difference equation~(\ref{diff eq_Bojo+CC}) as an 
evolution equation with respect to the discrete time.
We can see that $\Psi(0)$ disappears for $\mu=0$ and $\pm 4\mu_{0} $. For this reason,  
the solution can be uniquely extended beyond the classical singularity $\mu=0$.
That is, given some two initial data $\Psi(\epsilon+4N\mu_{0})$  and $\Psi(\epsilon+4(N+1)\mu_{0} )$ 
for $\epsilon\in (0,4\mu_{0})$ and a natural number $N$, 
one can uniquely determine the values $\Psi(\epsilon+4n \mu_{0} )$ for $n=0,\pm 1, \pm 2, \cdots$. 
For the $\epsilon=0$ case,
given some two initial data $\Psi(4N\mu_{0})$ and $\Psi(4(N+1)\mu_{0} )$,
the difference equation (\ref{diff eq_Bojo+CC}) generates $\Psi(4n\mu_{0})$ for $n=1,2,\cdots$
but the set of $\Psi(8\mu_{0})$ and $\Psi(4\mu_{0})$ does not generally satisfy 
Eq.~(\ref{diff eq_Bojo+CC}). This means that $\Psi(4N\mu_{0})$ and $\Psi(4(N+1)\mu_{0} )$
are constrained,
and once this is satisfied, we can uniquely 
determine the values $\Psi(4n \mu_{0} )$ for $n=\pm 1, \pm 2, \cdots$. 
Thus, we can conclude that the system has no initial singularity in this operator ordering. 
Although $\Psi(0)$ is left undetermined in Eq. (3.4), we can determine 
unambiguously the discrete evolution beyond $\mu=0$. That is, the unique 
quantum evolution is not affected in spite of the undetermined value at 
$\mu=0$, which was the initial singularity in classical theory. This ambiguity 
is fixed so that $\Psi(0)=0$ in Ref.~\cite{Bojowald4}  and we also use this value in our 
numerical analysis. 

Here we should note the superselection of the kinematical Hilbert space.
As seen from Eq.~(\ref{diff eq_Bojo+CC}),
a subspace $\mathcal{H}_{\epsilon}$ with a fixed parameter 
$\epsilon\in [0,4\mu_{0})$ 
spanned by a basis $\{|\mu\rangle : \mu = \epsilon + 4n\mu_0,  n \in \mathbb{Z} \}$ 
of the kinematical Hilbert space $\mathcal{H}_{\mathrm{kin}}^{\mathrm{grav}}$
are closed with respect to the action of the Hamiltonian constraint.
More precisely, the kinematical Hilbert space is naturally decomposed into
the sectors with respect to the action of the Hamiltonian constraint as 
\[
\mathcal{H}_{\mathrm{kin}}^{\mathrm{grav}}=\bigoplus_{\epsilon\in [0,4\mu_0
)} \mathcal{H}_{\epsilon}.\]
We refer to this decomposition as the superselection and to these subspaces as
the superselection sectors.
The superselection in LQC indicates that for each sector
we can fix the value of $\epsilon$ and work  
within this sector $\mathcal{H}_{\epsilon}$.
Since there is no initial singularity for the sectors with 
$\epsilon\ne 0$ in this case, we only have to choose the sector with $\epsilon = 0$
to study the presence or absence of the initial singularity at $\mu=0$.
On the other hand, when we consider the large $\mu$ limit, 
the value of $\epsilon$ does not affect the qualitative 
behaviour of $\Psi(\mu)$ and hence we can focus on the sector with $\epsilon=0$ again.
This is also the case in the equi-volume discretisation.

Next we consider the following operator ordering:
	\begin{eqnarray}\label{Hamiltonian_operator_pre2}
	 \widehat{C}_{\mathrm{grav}}=\frac{1}{16\pi l_{\mathrm{Pl}}^2\gamma^2}
         \left(\frac{96i\left(\mathrm{sgn}(p)\right)}{8\pi\gamma l^2_{\mathrm{Pl}}\mu_{0} ^{3}}
	 \widehat{EE}\widehat{F}+2\gamma^2\Lambda\widehat{V}\right).
	\end{eqnarray}
Then, the difference equation is given by 
        \begin{equation}\label{diff eq_Bojo+CC2}
          \left|-V_{\mu+\mu_{0} } + V_{\mu-\mu_{0} }\right| \bigl( \Psi(\mu+4\mu_{0} ) 
         - 2\Psi(\mu) + \Psi(\mu-4\mu_{0} ) \bigr) 
	 + \frac{16\pi\gamma^3 l^2_{\mathrm{Pl}}\mu_{0} ^{3}}{3}\Lambda V_{\mu} \Psi(\mu)= 0 . 
	\end{equation}
We can see that Eq.~(\ref{diff eq_Bojo+CC2}) becomes trivial for $\mu=0$ and 
hence we cannot determine $\Psi(-4\mu_{0} )$ from $\Psi(4\mu_{0} )$ and $\Psi(0)$. 
This fact indicates that we cannot determine all $\Psi(\mu)$ through $\mu=0$
from the data $\Psi(\mu)$ for $\mu>0$. 
In this sense,  
the model contains the initial singularity at $\mu=0$, beyond 
which the evolution cannot be uniquely extended.
This means that the absence of the singularity depends on the choice of
the operator ordering in the quantisation of the Hamiltonian. 

\subsection{Large volume limit problem }\label{sec:large_volume_limit}

In this section, we shall see the large-$\mu$ behaviour of the wave function $\Psi(\mu)$ 
determined by the difference equation and discuss its physical implication. 
Since the effects of the discreteness are dominant in the Planck scale physics but 
should disappear when the universe becomes large, 
in LQC the discrete wave function in the large volume is expected to be well approximated by 
a smooth wave function.
If this naive expectation is valid, it is natural to think that
the smooth wave function is described by a solution to the WDW equation
which is obtained by quantising the system in the usual Schr\"{o}dinger representation.
We give a brief derivation of the WDW equation in Appendix \ref{wdw eq}.

\begin{figure}[h!]
 \begin{center}
  \subfigure[]{
   \includegraphics[width=.45 \columnwidth]{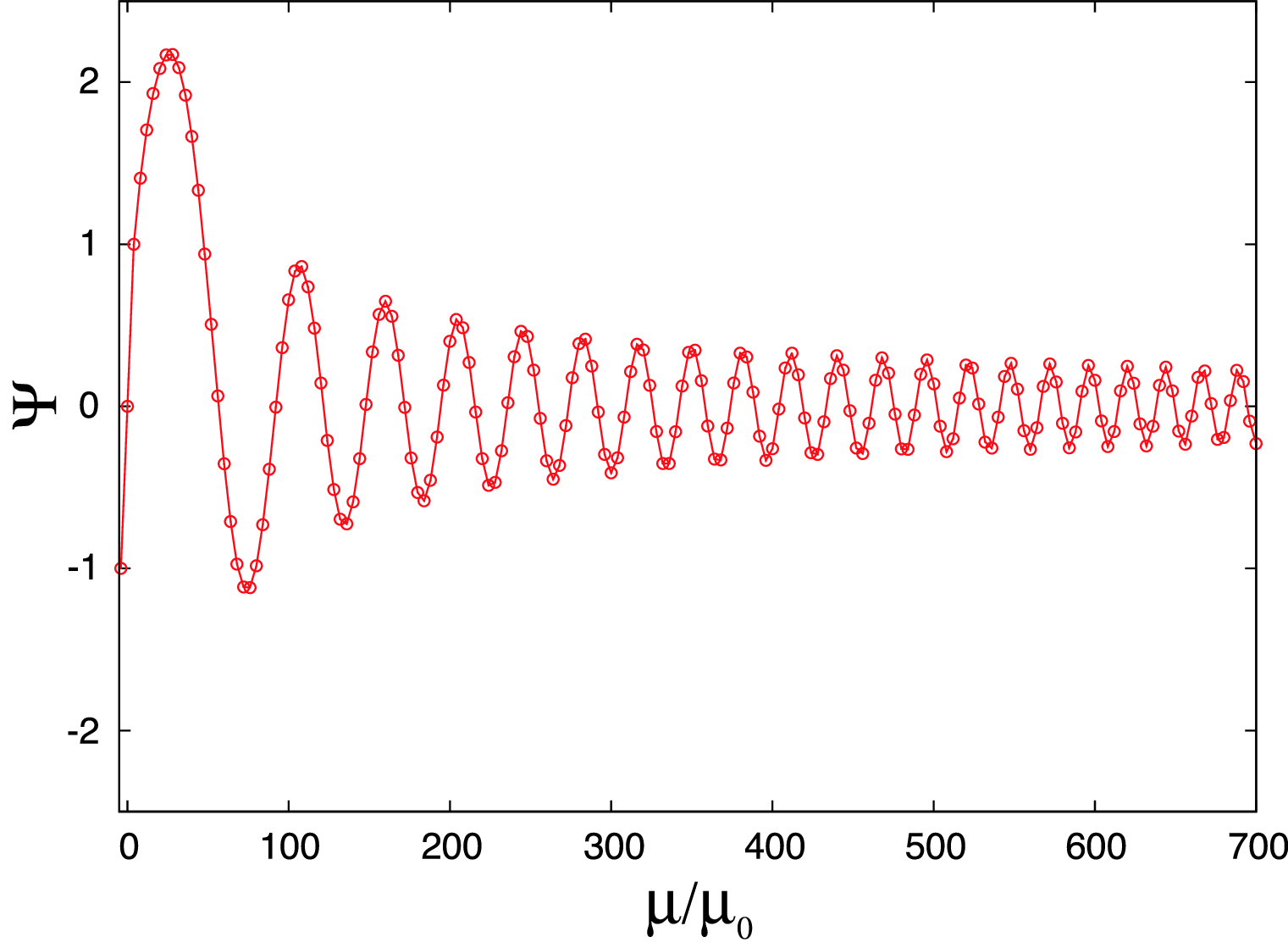}
  \label{fig:wave_area_1}}~
  \subfigure[]{
   \includegraphics[width=.45 \columnwidth]{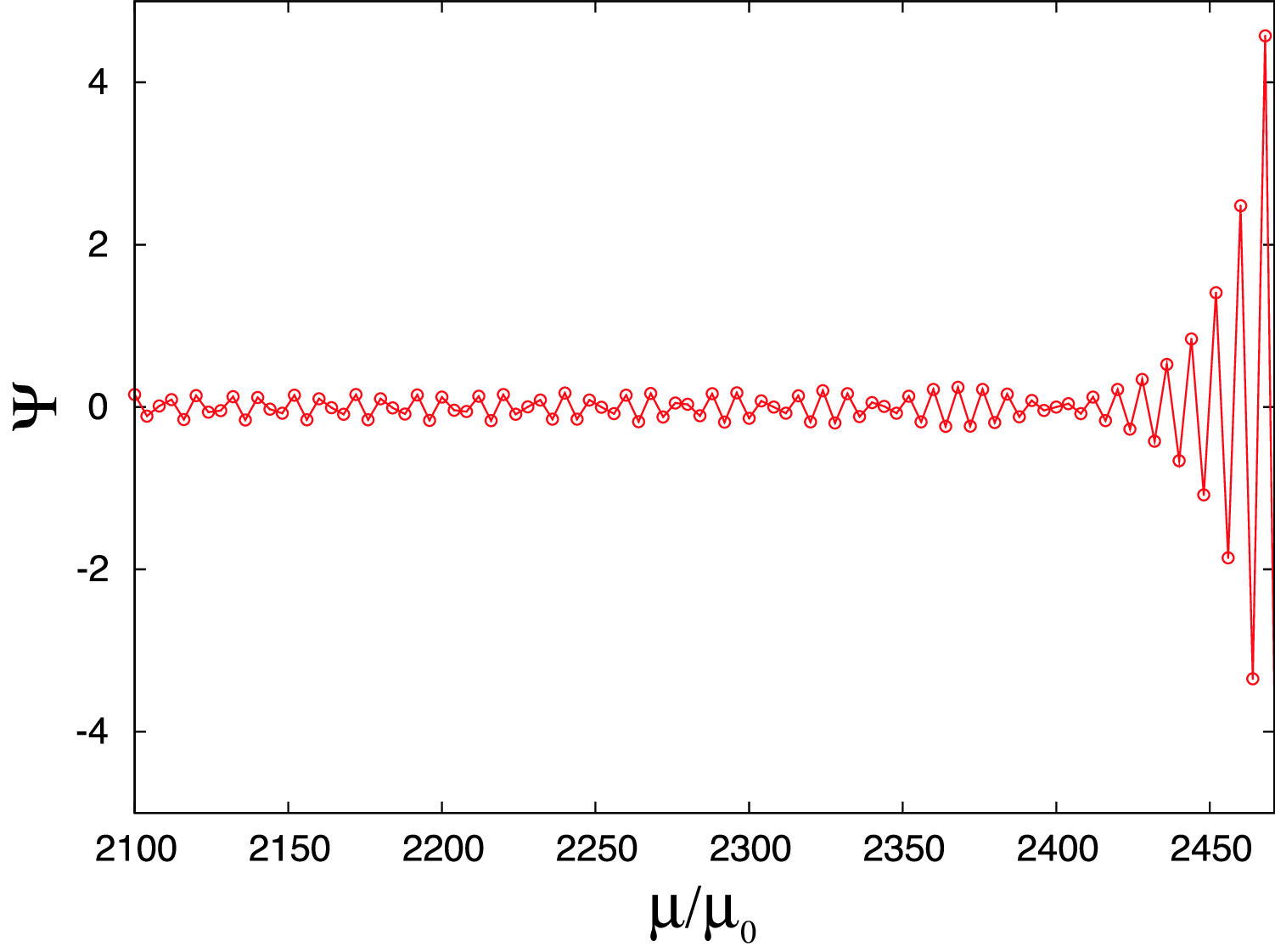}
  \label{fig:wave_area_2}}\\
  \subfigure[]{
   \includegraphics[width=.45 \columnwidth]{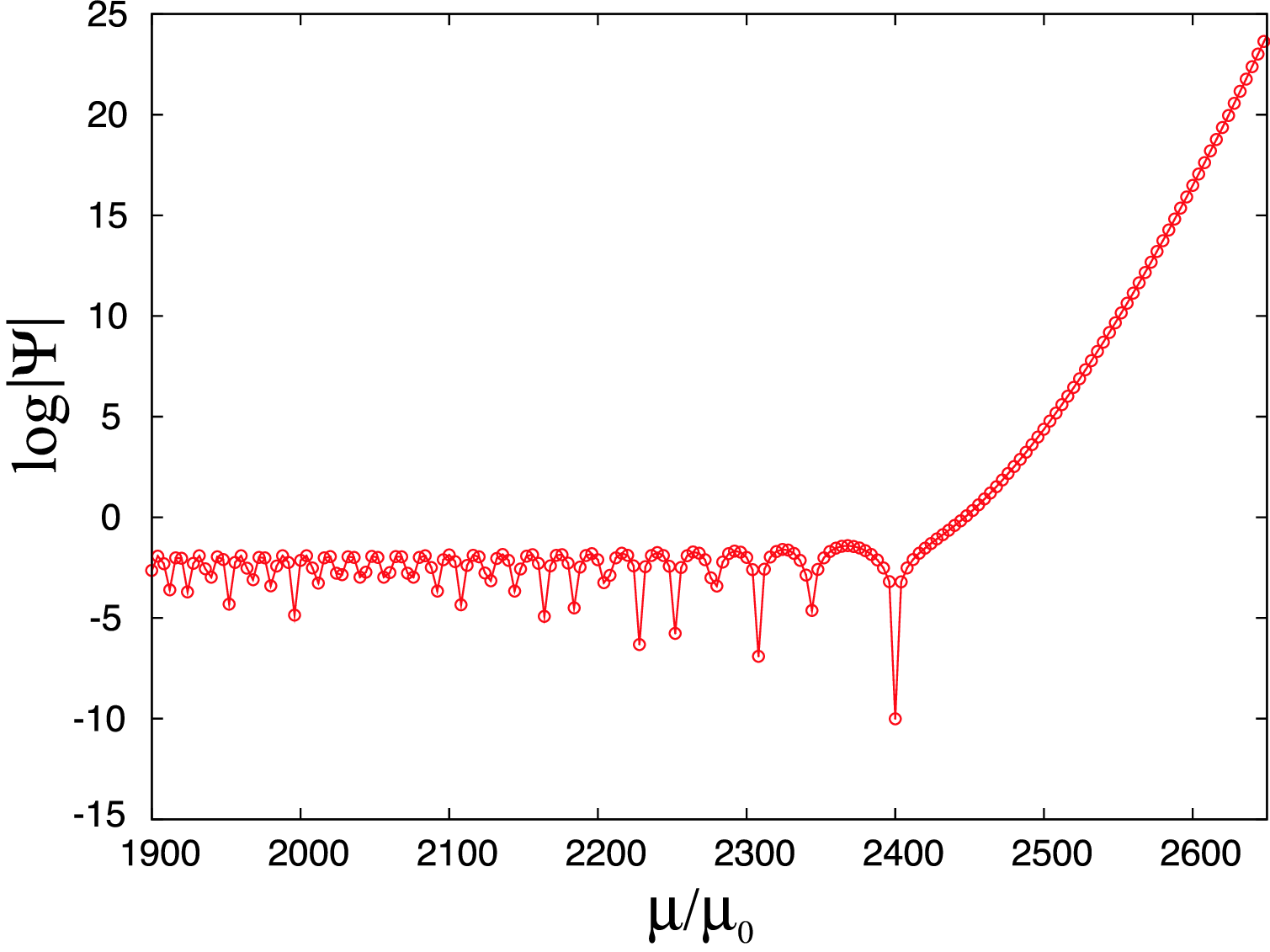}
  \label{fig:log}}~
  \caption{The wave function $\Psi(\mu)$ for (a) $0 \le \mu/\mu_0 \le 700$ and 
  	(b) $2100 \le \mu/\mu_0 \alt 2450$, and 
  	(c) the logarithmic scale of its absolute value are plotted as a function of $\mu$ with 
  	$\tilde{\Lambda}\mu_{0} ^3=0.005$, where we choose 
        the initial values as $\Psi(-4\mu_{0} )=-1$ and $\Psi(0)=0$.
        } 
  \label{fig:1}
 \end{center}
\end{figure}

To see the large-$\mu$ behaviour of the discretised wave function $\Psi(\mu)$, 
we numerically solve the difference equation (\ref{diff eq_Bojo+CC}).
We first choose the initial values $\Psi(-4\mu_{0} )$ and $\Psi(0)$,
and then determine $\Psi(4n\mu_{0} )$ for $n=1,2,...$ by the difference equation.
Here we define the dimensionless cosmological constant as $\tilde{\Lambda}=(16\pi/3)\gamma^3l^2_{\mathrm{Pl}}\Lambda$.
In the numerical calculation below, we set $\tilde{\Lambda}\mu_{0} ^3=0.005$ 
and these initial values as $\Psi(-4\mu_{0} )=-1$ and $\Psi(0)=0$,
where the values have no particular meaning 
and are selected to make the plots of the wave function visible.
Figs.~\ref{fig:wave_area_1} and \ref{fig:wave_area_2} show the wave function 
$\Psi(\mu)$ as a function of $\mu=4n\mu_{0} $,
and Fig.~\ref{fig:log} is for the logarithmic 
scale of $|\Psi(\mu)|$. 
As Fig.~\ref{fig:wave_area_1} illustrates,
the wave function $\Psi(\mu)$ can be regarded as the sampling of 
a decaying sinusoidal oscillation at the points sufficiently near the origin.
However, as seen from Fig.~\ref{fig:wave_area_2},
$\Psi(\mu)$ drastically changes its behaviour at $\mu \simeq 2400 \mu_0$,
so that $\Psi(\mu)$ flips its sign 
at each step and its absolute value grows up very rapidly for $\mu\agt 2400 \mu_{0} $.
We can see from Fig.~\ref{fig:log} that $|\Psi(\mu)|$ grows approximately
exponentially for $\mu\agt 2400\mu_{0}$.

This behaviour can be deduced from the following simple argument.
We have the recursive relation (\ref{diff eq_Bojo+CC}).
For $\mu\gg \mu_{0} $, we have
	\begin{equation}
	 \Psi(\mu+4\mu_{0} )-\left(2- \frac{1}{3}\mu\mu_{0} ^{2}\tilde{\Lambda}\right)\Psi(\mu)
	 +\Psi(\mu-4\mu_{0} )=0.
	 \label{eq:recursive_equiarea_largemu}
	\end{equation}
If $ |\mu\mu_{0} ^{2}\tilde{\Lambda}/6|\ll 1$, it is easily found that the solution is given by 
	\begin{equation}
	 \Psi(\mu_{0}+4n\mu_{0} ) \approx A n+B, 
	\end{equation}
where $A$ and $B$ are constants. However, if $ |\mu\mu_{0} ^{2}\tilde{\Lambda}/6| \gg 1$, 
in the present setting, we generally get a solution for which the last term on the left hand side of Eq.~(\ref{eq:recursive_equiarea_largemu}) can be neglected, i.e., 
	\begin{equation}
	 \Psi(\mu+4\mu_{0})\approx \left(2- \frac{1}{3}\mu\mu_{0} ^{2}\tilde{\Lambda}\right)
	 \Psi(\mu).
	\end{equation}
For $ |\mu\mu_{0} ^{2}\tilde{\Lambda}/6|\gg 1$, the factor $ (2- \mu\mu_{0} ^{2}\tilde{\Lambda}/3)$ 
is negative and positive if the cosmological constant is positive and negative,
respectively.  In both cases, its absolute value is
much larger than unity from the assumption. Thus, 
$\Psi(\mu)$ increases its absolute value at each step 
by the factor which is much larger than unity and 
the growth is approximately exponential. 
It is clear that the latter case cannot be regarded as smooth 
in spite of the very large volume.
We can expect that this peculiar behaviour becomes prominent for 
$\mu\simeq 9 \mu_{0} ^{-2}\tilde{\Lambda}^{-1}$ for $\tilde{\Lambda}>0$ and this is 
consistent with the numerical result plotted in Fig.~\ref{fig:1}. 

On the other hand, for $\mu \gg \mu_0$, 
the difference equation (\ref{diff eq_Bojo+CC}) can be approximated by the following WDW
equation under the assumption that the wave function varies sufficiently slowly (see Appendix~\ref{continuum limit} for a detailed derivation): 
	\begin{align}
	 \frac{\mathrm{d}^2}{\mathrm{d}\mu^2} \left(\sqrt{\mu}\Psi(\mu)\right) + \frac{\pi\gamma^3l^2_{\mathrm{Pl}}}{9}\mu^{3/2}\Lambda\Psi(\mu) = 0.\label{wdw} 
	\end{align}	
The general solution of the differential equation (\ref{wdw}) for $\mu \ge 0$ takes the form
	\begin{align}
	 \Psi(\mu)=
	 \mu^{-\frac{1}{2}}\left[ C_1 \textrm{Ai} \left( -\alpha_1^{\frac{1}{3}}\mu\right) 
                 +C_2 \textrm{Bi} \left( -\alpha_1^{\frac{1}{3}}\mu\right)\right],
                 \label{eq:wdw_solution}
	\end{align}
where $C_1$ and $C_2$ are arbitrary constants, 
$\mathrm{Ai}$ and $\mathrm{Bi}$ are the Airy functions 
and we have defined $\alpha_1:=(\pi/9)\gamma^3l_{\mathrm{Pl}}^2\Lambda$.
We present the WDW equations and its general solutions for some
operator orderings in Appendix~{\ref{wdw eq}}.
In the large $\alpha_1^{\frac{1}{3}}\mu$, the above 
function approaches a decaying sinusoidal curve.
Thus, the behaviour of the solution, which is 
plotted in Fig.~\ref{fig:1}, to the difference 
equation (\ref{diff eq_Bojo+CC}) is quite different from 
that of the solution (\ref{eq:wdw_solution}) 
to the differential equation (\ref{wdw}) 
even when the volume of the universe becomes large.
In other words, in the equi-area discretisation, the solution of the difference equation 
cannot be regarded as a smooth wave function for large $\mu$
in the presence of the cosmological constant.
We refer to this problem as {\it the large-volume limit problem}.

\section{Equi-volume discretisation} \label{sec:equi-volume}

\subsection{Resolution of the large volume limit problem}
In this section, we shall consider the large volume limit in the 
equi-volume discretisation.  We should note that the Hamiltonian constraint is written as 
	\begin{equation}\label{H_Ash}
	 {C}_{\mathrm{grav}} = \frac{1}{16\pi l_{\mathrm{Pl}}^2\gamma^2}
         \left(\frac{96i (\mathrm{sgn}(p))}{8\pi\gamma l^2_{\mathrm{Pl}}}\widehat{\frac{1}{\bar{\mu}^3}}
         \widehat{{\sin}^2\frac{\bar{\mu}c}{2}{\cos}^2\frac{\bar{\mu}c}{2}} 
	 \widehat{\left[{\sin}\left(\frac{\bar{\mu} c}{2}\right){V}{\cos}\left(\frac{\bar{\mu} c}{2}\right)-
         {\cos}\left(\frac{\bar{\mu} c}{2}\right){V}{\sin}\left(\frac{\bar{\mu} c}{2}\right)\right]}
         +2\gamma^2 \Lambda \widehat{V}\right),
	\end{equation}
where the ambiguity in the operator ordering is neglected.
Since $\bar{\mu}$ is a function of $p$, $1/\bar{\mu}^3$ becomes also an 
operator in the quantisation.
We put 
	\begin{equation}
	 \widehat{F}=\widehat{{\sin}^2\frac{\bar{\mu}{c}}{2}{\cos}^2\frac{\bar{\mu}c}{2}}
	 , \quad 
	 \widehat{EE}=\widehat{{\sin}\left(\frac{{\bar{\mu}} {c}}{2}\right){V}{\cos}\left(\frac{{\bar{\mu}}{c}}{2}\right)-
	 {\cos}\left(\frac{{\bar{\mu}} {c}}{2}\right){V}{\sin}\left(\frac{{\bar{\mu}} {c}}{2}\right)}.
	\end{equation} 
We first choose the following operator ordering:
	\begin{equation}
	 {C}_{\mathrm{grav}}=\frac{1}{16\pi l_{\mathrm{Pl}}^2\gamma^2}
         \left(\frac{96i(\mathrm{sgn}(p))}{8\pi\gamma l^2_{\mathrm{Pl}}}\widehat{F}
	 \widehat{\frac{1}{\bar{\mu}^{3}}}\widehat{EE}+2\gamma^{2}\Lambda\widehat{V}\right),
	\end{equation}
and we will see the other choices in the next section. 

\begin{figure}[h!]
 \begin{center}
  \subfigure[]{
   \includegraphics[width=.45 \columnwidth]{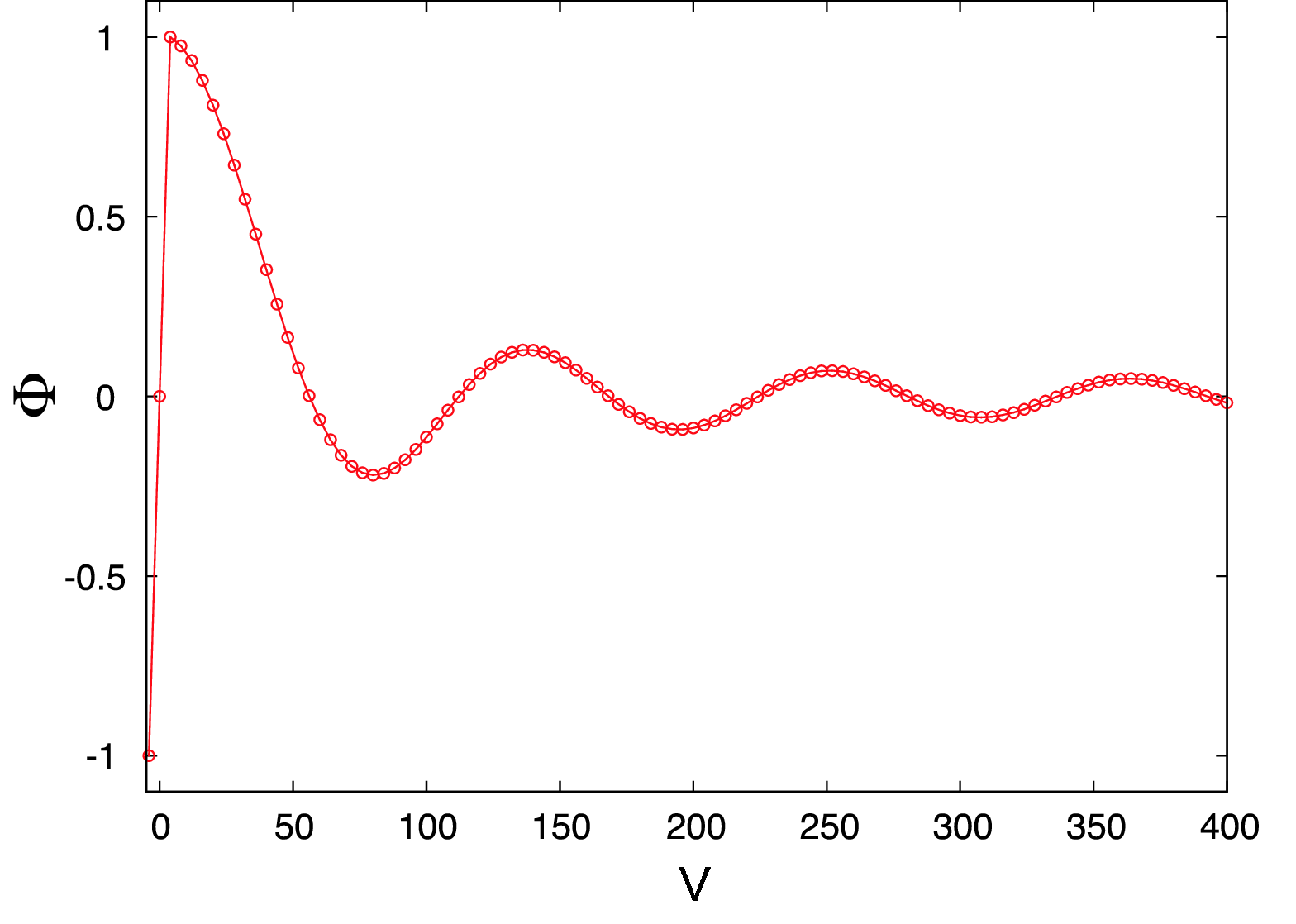}
  \label{fig:wave_vol_1}}~
  \subfigure[]{
   \includegraphics[width=.45 \columnwidth]{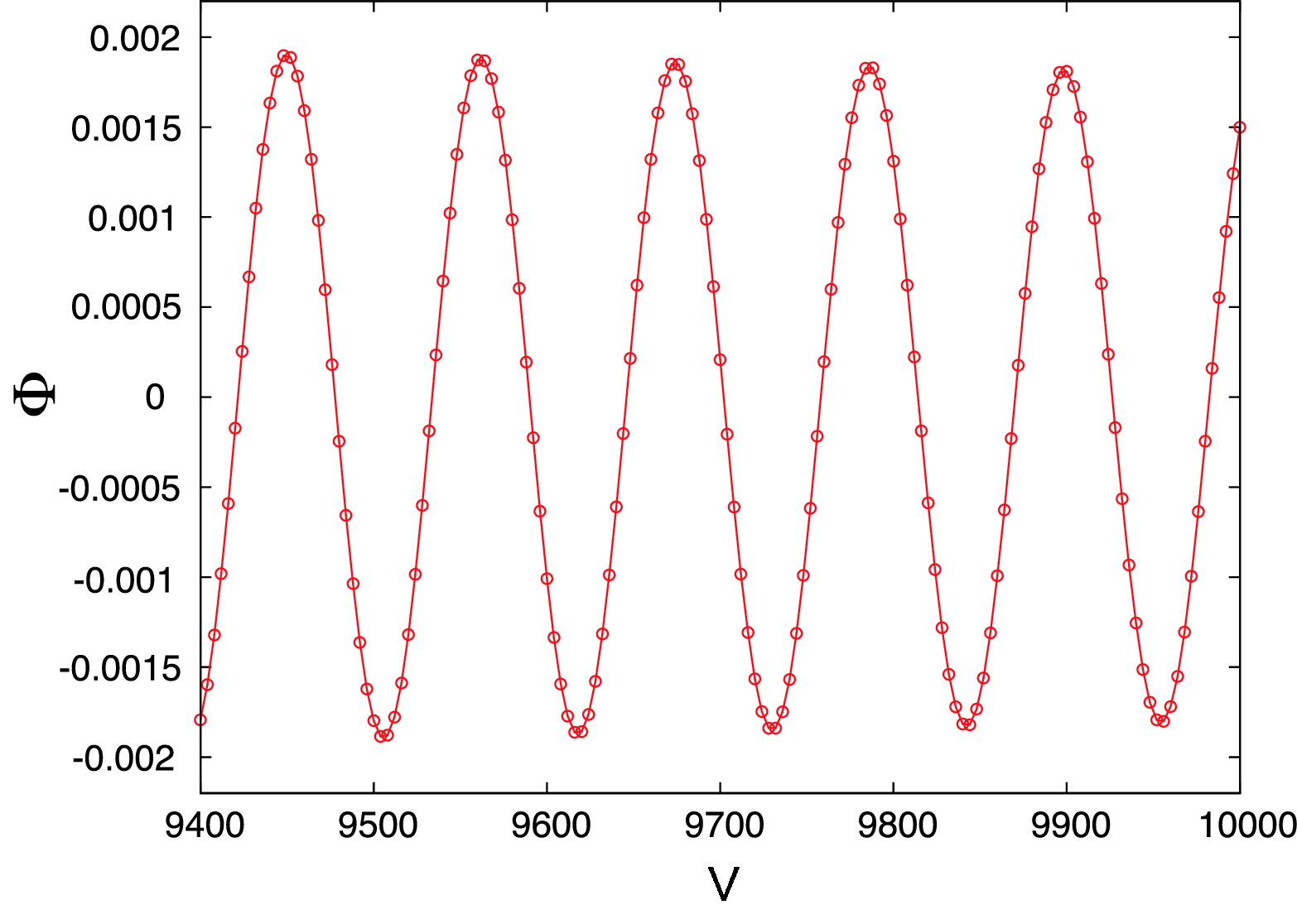}
  \label{fig:wave_vol_2}}
    \caption{The wave function $\Phi(v)$ for (a) $0 \le v \le 400$ and 
    (b) $9400 \le v \le 10000$ are plotted as a function of 
$v=4n$ with 
	 $\sqrt{3}\tilde{\Lambda}=0.1$, where we choose the initial values as $\Phi(-4)=-1$ and $\Phi(0)=0$.}
	 \label{fig:2}
 \end{center}
\end{figure}

As we have seen in Sec.\ref{subsec:discretisation ambiguity}, in this case, it is convenient to use a new label $v$ instead of $\mu$.
We rewrite $\widehat{1/\bar{\mu}^3}$ in terms of $\widehat{V}$ 
as 
	\begin{equation}\label{eq:mubar}
	\widehat{\frac{1}{\bar{\mu}^{3}}} = \left(\frac{6}{8\pi\gamma l_{\mathrm{Pl}}^{2}}\right)^{3/2}\frac{K\widehat{V}}{\sqrt{3}}, 
	\end{equation}
by using Eqs.~(\ref{eq:phatmu}), (\ref{barmu}) and (\ref{eq:vmu}).
Then, the Hamiltonian constraint yields the following difference equation:  
	\begin{eqnarray}\label{diff eq_Ash+CC}
	 && |v+4| \left||v+3|-|v+5|\right| \Phi(v+4) \nonumber \\
	 && \quad - \left( 2|v| \left||v-1|-|v+1|\right| - 
         \frac{16\sqrt{3}\pi}{3}\gamma^3 l^2_{\mathrm{Pl}}\Lambda |v| \right) \Phi(v)
	 + |v-4| \left||v-5|-|v-3|\right| \Phi(v-4) 
	 = 0, \label{diff_eq v}
	\end{eqnarray}
where $\Phi(v)=\langle v | \Phi \rangle$. 
We numerically obtain the solution of the above difference equation with 
$\sqrt{3}\tilde{\Lambda}=0.1$
from the initial values $\Phi(-4)=-1$ and $\Phi(0)=0$.  
Again, we have chosen these values to make the plots visible. 
The obtained wave function $\Phi(v)$ is plotted 
in Figs.~\ref{fig:wave_vol_1} and \ref{fig:wave_vol_2}. 
We see from these figures that the discretised wave function oscillates and can be regarded as 
the sampling of a smooth sinusoidal curve for large $v$. 
This means that the large volume limit 
problem in the equi-area discretisation is resolved in the equi-volume discretisation. 
Indeed, as presented in Appendix~\ref{continuum limit}, 
the difference equation~(\ref{diff_eq v}) can be approximated by the 
WDW equation
	\begin{align}
	 \frac{\mathrm{d}^2}{\mathrm{d} v^2} \bigl(|v| \Phi(v)\bigr)
 	 + \frac{4\pi}{81K^2}\gamma^3 l^2_{\mathrm{Pl}}\Lambda |v|\Phi(v) = 0,
	\end{align}
and the general solution is written as
	\begin{align}
	 \Phi(v)=\frac{1}{v}\left(C_1 e^{i\sqrt{\alpha_2} v} + C_2 e^{-i\sqrt{\alpha_2} v}\right),
         \label{wdw_v}
	\end{align}
where $C_1$ and $C_2$ are arbitrary constants and we have defined $\alpha_2=\frac{4\pi}{81K^2}\gamma^3 l^2_{\mathrm{Pl}}\Lambda$.  
The wave function plotted in Fig.~\ref{fig:2} has good agreement with
the solution (\ref{wdw_v}) for large $v$. 

We can also analytically show that the large volume limit should be resolved in this 
discretisation. 
For $v\gg 1$, we can approximate Eq.~(\ref{diff eq_Ash+CC}) by 
	\begin{equation}
	 \Phi(v+4)-2(1-C\tilde{\Lambda})\Phi(v)+\Phi(v-4)=0,
	\end{equation}
where $C=2/(27K^{2})$. The characteristic equation for this 
recursive relation is the following:
	\begin{equation} 
	 x^{2}-2(1-C\tilde{\Lambda})x+1=0.
	\end{equation}
The roots are given by 
	\begin{equation}
	 x=(1-C\tilde{\Lambda})\pm \sqrt{(1-C\tilde{\Lambda})^{2}-1}.
	\end{equation}
From the above, we can deduce the large-$v$ behaviour of $\Phi(v)$.
For $0<C\tilde{\Lambda}<2$, $\Phi(4n)$ purely oscillates 
sinusoidally. 
For $C\tilde{\Lambda}<0$ or $C\tilde{\Lambda}>2$, it is generally 
dominated by the exponential growth. For $C\tilde{\Lambda}=0$, 
$\Phi(4n)=An+B$, where $A$ and $B$ are constants. For $C\tilde{\Lambda}=2$, 
$\Phi(4n+4)+\Phi(4n)=(-1)^{n}D$, where $D$ is a constant.
Thus, if $|C\tilde{\Lambda}|\ll 1$, which is realistic in our Universe,
$\Phi(4n)$ changes very slowly at each step whether it oscillates with the angular velocity 
$\sim \sqrt{C\tilde{\Lambda}}$ or grows exponentially with the growth rate
$\sim \sqrt{C\tilde{\Lambda}}$. This means that we can physically regard $\Phi(v)$
as a smooth wave function in the large volume limit.

\subsection{Absence of singularity and operator ordering}

In this section, we shall discuss the initial singularity in the equi-volume discretisation.
We here demonstrate the following four typical types of the operator ordering in the Hamiltonian 
constraint in the equi-volume discretisation and show that the existence or absence of the initial singularity depends on the choice of the 
operator ordering.
	\begin{subequations}
	\begin{eqnarray}
	\widehat{C}_{\mathrm{grav}}^{(a)}&=&\frac{1}{16\pi l_{\mathrm{Pl}}^2\gamma^2}
        \left(\frac{96i(\mathrm{sgn}(p))}
	{8\pi\gamma l^2_{\mathrm{Pl}}}\widehat{F}\widehat{\frac{1}{\bar{\mu}^{3}}}\widehat{EE}+2\gamma^{2} \Lambda\widehat{V}\right), \\
	\widehat{C}_{\mathrm{grav}}^{(b)}&=&\frac{1}{16\pi l_{\mathrm{Pl}}^2\gamma^2}
        \left(\frac{96i(\mathrm{sgn}(p))}
	{8\pi\gamma l^2_{\mathrm{Pl}}}\widehat{\frac{1}{\bar{\mu}^{3}}}\widehat{F}\widehat{EE}+2\gamma^{2} \Lambda\widehat{V}\right), \\
	\widehat{C}_{\mathrm{grav}}^{(c)}&=&\frac{1}{16\pi l_{\mathrm{Pl}}^2\gamma^2}
        \left(\frac{96i(\mathrm{sgn}(p))}
	{8\pi\gamma l^2_{\mathrm{Pl}}}\widehat{EE}\widehat{\frac{1}{\bar{\mu}^{3}}}\widehat{F}+2\gamma^{2} \Lambda\widehat{V}\right), \\
	\widehat{C}_{\mathrm{grav}}^{(d)}&=&\frac{1}{16\pi l_{\mathrm{Pl}}^2\gamma^2}
        \left(\frac{96i(\mathrm{sgn}(p))}
	{8\pi\gamma l^2_{\mathrm{Pl}}}\widehat{EE}\widehat{F}\widehat{\frac{1}{\bar{\mu}^{3}}}+2\gamma^{2} \Lambda\widehat{V}\right). 
	\end{eqnarray}
	\end{subequations}
We call the above four types of the orderings (a), (b), (c) and (d).
Note that $\widehat{EE}$ and $\widehat{1/\bar{\mu}^3}$ are commutative 
with each other. 
We can also take different orderings. 
For example, 
$\widehat{1/\bar{\mu}^3}$ could be divided into two and put on both sides of $\widehat{F}$. 
We will discuss these orderings later.

The above Hamiltonian constraints for orderings (a) - (d) yield respectively 
the following difference equations:
\begin{subequations} 
	\begin{eqnarray}
		 &&  |v+4| \left||v+3|-|v+5|\right| \Phi(v+4)  \nonumber \\
		&& - \left( 2|v| \left||v-1|-|v+1|\right| - \frac{128\pi}{81}\gamma^3\frac{l^2_{\mathrm{Pl}}}{K^2}\Lambda |v| \right) \Phi(v)  + |v-4| \left||v-5|-|v-3|\right| \Phi(v-4) = 0 \label{type a},  \\
		 &&  |v| \left||v+5|-|v+3|\right| \Phi(v+4) \nonumber \\
		&& - \left( 2|v| \left||v+1|-|v-1|\right| - \frac{128\pi}{81}\gamma^3\frac{l^2_{\mathrm{Pl}}}{K^2}\Lambda |v| \right) \Phi(v) + |v| \left||v-3|-|v-5|\right| \Phi(v-4) =0, \label{type b} \\
		 &&  |v| \left||v+1|-|v-1|\right| \Phi(v+4)  \nonumber \\
	 	&& - \left( 2|v| \left||v+1|-|v-1|\right| - \frac{128\pi}{81}\gamma^3\frac{l^2_{\mathrm{Pl}}}{K^2}\Lambda |v| \right) \Phi(v) + |v| \left||v+1|-|v-1|\right| \Phi(v-4) =0,\label{type c} \\
		 && |v+4| \left||v+1|-|v-1|\right| \Phi(v+4)  \nonumber \\
		&& - \left( 2|v| \left||v+1|-|v-1|\right| - \frac{128\pi}{81}\gamma^3\frac{l^2_{\mathrm{Pl}}}{K^2}\Lambda |v| \right) \Phi(v) + |v-4| \left||v+1|-|v-1|\right| \Phi(v-4) =0.\label{type d}
	\end{eqnarray}
\end{subequations}
The WDW equations obtained by the approximation of the difference equations
(\ref{type a})-(\ref{type d})  are presented in Appendix~\ref{continuum limit}.
We can see that $\Phi(0)$ disappears in any of these equations. However, only in Eq.~(\ref{type a}),
$\Phi(4)$ can be directly related to $\Phi(-4)$. This leads to the following conclusion. 
Suppose that we are given $\Phi(v)$ for all $v>0$. 
Then, only ordering (a) among the above four choices 
determines $\Phi(v)$ for all $v<0$. 
For the other choices, the difference equation does not determine
$\Phi(4n)$ for $n=-1,-2,-3,...$. In this sense, the absence of the singularity is possible
in and only in ordering (a) among the choices (a)-(d). 
Thus, among the six models demonstrated in the present paper,  
the model with the equi-volume discretisation and ordering (a) is the only model where the initial singularity is 
absent and has the large volume limit as a smooth wave function.
In ordering (a), if only we fix $\Phi(4)$, we can determine $\Phi(4n)$ for $n=\pm 1, \pm 2, \cdots$. 
As in the equi-area discretisation, although $\Phi(0)$ is left undetermined, 
this does not affect the unique quantum evolution beyond $v=0$.

Moreover, we can find a general ordering rule required for the absence of the singularity : 
If the volume operator $\widehat{V}$ or its positive power, 
which is included in $\widehat{EE}$ and $\widehat{1/\bar{\mu}^3}$, 
appears in front of $\widehat{F}$, 
the initial singularity appears. 
We can understand this fact as follows.
For simplicity, we here assume that there is a constraint equation as $C = FV=0$ classically.
In quantum theory, there are two choices of the operator ordering for the constraint, and then 
the action of the constraint operators $\widehat{C_1}=\widehat{V}\widehat{F}$ and $\widehat{C_2}=\widehat{F}\widehat{V}$ 
on the state $\vert v \rangle$ is given by 
	\begin{eqnarray}
	 \widehat{C_1}\vert v\rangle&=&\widehat{V}\widehat{F}\vert v\rangle= -\frac{1}{16}\bigl(V_{v+4}\vert v+4 \rangle-2V_{v}\vert v\rangle+V_{v-4}\vert v-4\rangle\bigr),\\
         \widehat{C_2}\vert v\rangle&=&\widehat{F}\widehat{V}\vert v\rangle= -\frac{V_v}{16}\bigl(\vert v+4\rangle-2\vert v\rangle+\vert v-4\rangle\bigr). 
	\end{eqnarray}
Then, the constraint equations for $\Phi(v)$ become 
	\begin{equation}\label{vf}
	 V_v\bigl(\Phi(v+4)-2\Phi(v)+\Phi(v-4)\bigr)=0, 
	\end{equation}
for $\widehat{C_1}$, and 
	\begin{equation}\label{fv}
	 V_{v+4}\Phi(v+4)-2V_{v}\Phi(v)+V_{v-4}\Phi(v-4)=0,
	\end{equation}
for $\widehat{C_2}$, respectively. 
In Eq.~(\ref{vf}), there is the same eigenvalue $V_v$ of the volume operator in each term, 
and then all coefficients vanish for $v=0$. 
For this reason, in this case, we cannot determine $\Phi(-4)$ from $\Phi(4)$ and $\Phi(0)$, 
and do not have the unique evolution. 
On the other hand, in Eq.~(\ref{fv}), the left hand side does not disappear even for $v=0$. 
Thus, we can determine the unique solution beyond $v=0$, and therefore there is no singularity. 

\section{Discussion}\label{sec:discussion}
To make the present analysis complete we also need to discuss the matter Hamiltonian $C_{\mathrm{matter}}$. 
For simplicity, we first focus on the equi-volume discretisation.
Matter fields can be incorporated by introducing the dependence of the wave function 
on the matter fields such as $\Phi(v,\phi)$. 
Then, we can see that 
the presence of the matter fields does not greatly change the discussion. 
For example, suppose we choose Eq.~(\ref{type a}) 
as the gravitational part of the difference equation.
Then, 
for $\Phi(v,\phi)$ we obtain the following evolution equation 
 \begin{eqnarray}\label{eq:matter}
         &&  |v+4| \left||v+3|-|v+5|\right| \Phi(v+4,\phi)- \left( 2|v| \left||v-1|-|v+1|\right| - \frac{128\pi}{81}\gamma^3\frac{l^2_{\mathrm{Pl}}}{K^2}\Lambda |v| \right) \Phi(v,\phi) \nonumber \\
  &&  ~~~~+ |v-4| \left||v-5|-|v-3|\right| \Phi(v-4,\phi) = -48Kl_{\mathrm{Pl}}\sqrt{\pi \gamma^3}\hat{C}_{\mathrm{matter}}\Phi(v,\phi), 
        \end{eqnarray}
where we have used the constraint equation $C_{\mathrm{grav}}+C_{\mathrm{matter}}=0$. 
As is explicitly shown in Appendix~\ref{sec:matter_hamiltonian}, 
for general matter fields with an exceptional case, 
the right hand side of Eq.~(\ref{eq:matter}) is proportional to $\Phi(v,\phi)$
and the coefficient vanishes for $v=0$.
This means that the presence or absence of 
the singularity does not depend on the matter fields in general.
It depends on the operator ordering of the gravitational 
Hamiltonian constraint. This is also true for the equi-area discretisation.

Whether the wave function recovers the smoothness 
in the large volume limit strongly depends on the choice of the discretisation as well as 
the form of the matter fields. 
Actually, this problem is already studied well in detail 
by Nelson and Sakellariadou~\cite{Nelson_Sakellariadou_2007a,Nelson_Sakellariadou_2007b}.
In this context, the present paper has shown that if we only 
include the cosmological constant, the large volume limit is 
problematic in the equi-area discretisation but 
this problem is resolved in the equi-volume discretisation.
In reality, the cosmological constant model with each discretisation falls into some particular set of 
the parameter values in their notation and we can find that 
clearly our result is consistent with theirs. 
On the other hand, we have generated the solution of the finite difference 
equation for both discretisations and shown how problematic the 
equi-area discretisation is and 
how nicely the equi-volume discretisation sorts this out.
Since this behaviour can be explained completely by taking the 
large volume limit of the discretised Hamiltonian constraint
equations and the choice of the operator ordering does not affect this limit, 
it is clear that the large volume limit is insensitive to the 
choice of the operator ordering.

\section{Conclusion}
\label{sec:conclusion}
We have investigated a homogeneous, isotropic and spatially flat 
universe with the cosmological constant 
in the context of LQC. In particular, we have studied theoretical ambiguities in 
the quantisation of the Hamiltonian, which arise in the operator ordering
and the discretisation and are hard to fix within the framework of LQC. 
We focus on two important features of LQC, the absence of the initial singularity and the large volume limit. 
We have shown that the absence of the initial singularity strongly depends on the choice of the 
operator ordering. Therefore, the requirement for the absence of the singularity can potentially 
make the ordering ambiguity very small. 
Furthermore, we have found a general ordering rule required for the absence of the singularity. 
On the other hand, the choice of the discretisation 
is crucial when we consider the large volume limit. We have demonstrated two typical choices of the discretisation, 
the equi-area and equi-volume ones. Then, we have shown that in the former case there arises a 
serious problem in the physical interpretation of the wave function as a continuous function 
in the large volume limit and also that this problem is resolved in the latter case. 
It is clear that we cannot fix these ambiguities only from the present results. However, if we can 
deduce the physical implications of full LQG even qualitatively, we will be able to fix 
the ambiguities of LQC as the phenomenological realisation of full LQG in the minisuperspace model. 

\section*{Acknowledgment}
We are very grateful to M. Bojowald and R. Tavakol for fruitful discussion and comments. We would also
like to thank A. Hosoya, K.-i. Maeda, H. Kodama, J. Soda, M. Siino
and S. Kamata for helpful comments. 
MS is supported by Rikkyo University Special Fund for Research.
TH was partly supported by the Grant-in-Aid for Scientific
Research Fund of the Ministry of Education, Culture,
Sports, Science and Technology, Japan [Young Scientists
(B) 18740144 and 21740190]. This work has benefited from exchange 
visits supported by a JSPS and Royal Society bilateral grant.
\appendix
\section{Wheeler-De Witt equation and its general solution \label{wdw eq}}

Here, we summarise the WDW theory for the flat FRW universe with the cosmological constant.
In terms of the connection-triad variables $c$ and $p$, the Poisson bracket takes the form
	\begin{align}
	 \{ c, p \}=\frac{8\pi G \gamma}{3},\label{pb}
	\end{align}
and the Hamiltonian constraint is written as
	\begin{equation}
	 C_{\mathrm{grav}}=\frac{1}{16\pi\gamma^2 G}\left(-6c^2\sqrt{|p|}+2\gamma^2\Lambda |p|^{\frac{3}{2}}\right)=0. \label{cp_eq}
	\end{equation}

To quantise the system, Dirac's method is used in the WDW theory.
That is, the Poisson bracket (\ref{pb}) is replaced by the commutation relation between the operators
corresponding to the canonical pair:
	\begin{align}
	 [ \widehat{c}, \widehat{p} ]=  \frac{ 8i\pi l_{\mathrm{Pl}}^2 \gamma}{3},
	\end{align}
and the classical constraint (\ref{cp_eq}) is promoted to a constraint 
for physical quantum states: $\widehat{C}_{\mathrm{grav}}\Psi=0$.
This quantum constraint is called the WDW equation.
We here consider the following three types of the operator ordering for the Hamiltonian constraint:
	\begin{subequations}
	\begin{align}
	 \widehat{C}_{\mathrm{grav}}^{(1)} &= \frac{1}{16\pi \gamma^2l_{\mathrm{Pl}}^2}\left(-6\widehat{c}^2\sqrt{|\widehat{p}|}+2\gamma^2\Lambda|\widehat{p}|^{\frac{3}{2}}\right),\label{const1} \\
	 \widehat{C}_{\mathrm{grav}}^{(2)} &= \frac{1}{16\pi \gamma^2l_{\mathrm{Pl}}^2}\left(-6\sqrt{|\widehat{p}|}\widehat{c}^2+2\gamma^2\Lambda|\widehat{p}|^{\frac{3}{2}}\right),\label{const2} \\
	 \widehat{C}_{\mathrm{grav}}^{(3)} &= \frac{1}{16\pi \gamma^2l_{\mathrm{Pl}}^2}\left(-6\widehat{c}\sqrt{|\widehat{p}|}\widehat{c}+2\gamma^2\Lambda|\widehat{p}|^{\frac{3}{2}}\right).\label{const3}
	\end{align}
	\end{subequations}
If we take the ordinary Schr\"{o}dinger representation, where $\widehat{p}$ and $\widehat{c}$ respectively act as multiplication and differentiation:
	\begin{align}
	 \widehat{p}\Psi=p\Psi, \quad 
         \widehat{c}\Psi=\frac{8i\pi l_{\mathrm{Pl}}^2 \gamma}{3}\frac{\mathrm{d}\Psi}{\mathrm{d} p}, 
	\end{align}
the quantum constraints (\ref{const1})-(\ref{const3}) yield the corresponding WDW equations
	\begin{subequations}
	\begin{align}
	 \widehat{C}_{\mathrm{grav}}^{(1)}\Psi^{(1)} &= 
	 \frac{8\pi l_{\mathrm{Pl}}^2 }{3}\left(\sqrt{|p|}\frac{\mathrm{d}^2 \Psi^{(1)}}{\mathrm{d} p^2} 
         + \frac{\mathrm{sgn}(p)}{\sqrt{|p|}}\frac{\mathrm{d} \Psi^{(1)}}{\mathrm{d} p}-
         \frac{1}{4}|p|^{-\frac{3}{2}}\Psi^{(1)} \right) 
         +\frac{\Lambda}{8\pi l_{\mathrm{Pl}}^2}|p|^{\frac{3}{2}}\Psi^{(1)}=0,\label{WDW1}\\
	 \widehat{C}_{\mathrm{grav}}^{(2)}\Psi^{(2)} &=
	 \frac{8\pi l_{\mathrm{Pl}}^2}{3}\sqrt{|p|}\frac{\mathrm{d}^2 \Psi^{(2)}}{\mathrm{d} p^2}
	 +\frac{\Lambda}{8\pi l_{\mathrm{Pl}}^2}|p|^{\frac{3}{2}}\Psi^{(2)}=0,\label{WDW2}\\
	 \widehat{C}_{\mathrm{grav}}^{(3)}\Psi^{(3)} &=		 
	 \frac{8\pi l_{\mathrm{Pl}}^2 }{3}\left(\sqrt{|p|}\frac{\mathrm{d}^2 \Psi^{(3)}}{\mathrm{d} p^2}
         + \frac{\mathrm{sgn}(p)}{2\sqrt{|p|}}\frac{\mathrm{d} \Psi^{(3)}}{\mathrm{d} p} \right)
         +\frac{\Lambda}{8\pi l_{\mathrm{Pl}}^2}|p|^{\frac{3}{2}}\Psi^{(3)}=0.\label{WDW3}
	\end{align}
	\end{subequations}
The general solutions of the WDW equations (\ref{WDW1})-(\ref{WDW3}) for $p \ge 0$ are given by
	\begin{subequations}
	\begin{align}
	 \Psi^{(1)}&=
	 p^{-\frac{1}{2}}\left[ C_1 \textrm{Ai} \left( -\alpha_3^{\frac{1}{3}}p\right) 
	 +C_2 \textrm{Bi} \left( -\alpha_3^{\frac{1}{3}}p\right)\right], \\
	 \Psi^{(2)}&=
	 C_1 \textrm{Ai} \left( -\alpha_3^{\frac{1}{3}}p\right) +
         C_2 \textrm{Bi} \left( -\alpha_3^{\frac{1}{3}}p\right),\\
	 \Psi^{(3)}&=
	 p^{\frac{1}{4}} \left[C_1 \textrm{J}_{-\frac{1}{6}} 
         \left( \frac{2}{3}\sqrt{\alpha_3}p^{\frac{3}{2}}\right)
         +C_2 \textrm{J}_{\frac{1}{6}} 
         \left( \frac{2}{3}\sqrt{\alpha_3}p^{\frac{3}{2}}\right)\right],
	\end{align}
	\end{subequations}
where $C_1$ and $C_2$ are arbitrary constants, $\textrm{Ai}$ and $\textrm{Bi}$ are the Airy functions, 
$\textrm{J}_{\pm \frac{1}{6}}$ is the Bessel function, and we have defined 
$\alpha_3:=3\Lambda/(8\pi l_{\mathrm{Pl}}^2 )^2$.

\section{Large volume limit of the difference equations}\label{continuum limit} 
Here, we consider the large volume limit of $\langle\mu|\widehat{C}_{\mathrm{grav}}|\Psi\rangle$, 
and then we shall show that it corresponds to the WDW equation.
First, we consider the operator orderings (\ref{Hamiltonian_operator_pre}) and (\ref{Hamiltonian_operator_pre2})
in the equi-area discretisation. 
The operator ordering (\ref{Hamiltonian_operator_pre}) yields 
\begin{eqnarray}
\langle\mu|\widehat{C}_{\mathrm{grav}}|\Psi\rangle &=& 
\frac{1}{16\pi^2l^2_{\mathrm{Pl}}\gamma^2}\frac{3}{8\pi\gamma l^2_{\mathrm{Pl}}\mu_0^3} 
\biggl[ \left| V_{\mu+5\mu_{0} }-V_{\mu+3\mu_{0} }\right| \Psi(\mu+4\mu_{0} )  \nonumber\\
 &-& \left(2\left| V_{\mu+\mu_{0} } - V_{\mu-\mu_{0} }\right| -\frac{16\pi\gamma^3 l^2_{\mathrm{Pl}}\mu_{0} ^{3}}{3}\Lambda V_{\mu}\right) \Psi(\mu)
 + \left| V_{\mu-3\mu_{0} }-V_{\mu-5\mu_{0} }\right| \Psi(\mu-4\mu_{0} ) \biggr]. \label{diff eq_Bojo+CC in appendix}
\end{eqnarray}

For $\mu\gg\mu_0$, we can expand $(V_{\mu+L\mu_0}-V_{\mu+M\mu_0})$ 
around $\mu_0$ as follows:
	\begin{eqnarray}
	 V_{\mu+L\mu_0}-V_{\mu+M\mu_0} = \left(\frac{8\pi\gamma l^2_{\mathrm{Pl}}}{6}\right)^{3/2}\mu^{3/2}
	 \left\{\frac{3}{2}(L-M)\frac{\mu_0}{\mu} 
	 + \frac{3}{8}(L^2-M^2)\left(\frac{\mu_0}{\mu}\right)^2 \right. \nonumber \\
	 \left. - \frac{1}{16}(L^3-M^3)\left(\frac{\mu_0}{\mu}\right)^3 
	 + O\left(\left(\frac{\mu_0}{\mu}\right)^4\right) \right\}.  \label{V-V expand}
	\end{eqnarray}
Substituting Eq.~(\ref{V-V expand}) into Eq.~(\ref{diff eq_Bojo+CC in appendix}), we obtain 
	\begin{eqnarray}
	\langle\mu|\widehat{C}_{\mathrm{grav}}|\Psi\rangle = 
	 \frac{1}{16\pi^2l^2_{\mathrm{Pl}}\gamma^2}\frac{3}{8\pi\gamma l^2_{\mathrm{Pl}}\mu_0^3}\left(\frac{8\pi\gamma l^2_{\mathrm{Pl}}}{6}\right)^{3/2}\mu^{3/2} \left[ 
	 3\left(\frac{\mu_0}{\mu}\right)\left\{\Psi(\mu+4\mu_0)-2\Psi(\mu)+\Psi(\mu-4\mu_0)\right\} \right. \nonumber \\
	 +6\left(\frac{\mu_0}{\mu}\right)^2\left\{\Psi(\mu+4\mu_0)-\Psi(\mu-4\mu_0)\right\} 
	 - \frac{1}{8}\left(\frac{\mu_0}{\mu}\right)^3\left\{\Psi(\mu+4\mu_0)-2\Psi(\mu)+\Psi(\mu-4\mu_0)\right\} \nonumber \\ 
	 \left. -6\left(\frac{\mu_0}{\mu}\right)^3\left\{\Psi(\mu+4\mu_0)+\Psi(\mu-4\mu_0)\right\}
	 + \frac{16\pi\gamma^3l^2_{\mathrm{Pl}}\mu_0^3}{3}\Lambda\Psi(\mu) + O\left(\left(\frac{\mu_0}{\mu}\right)^4\right)
	 \right] .  \label{diff eq_Bojo+CC+V-V expand}
	\end{eqnarray}
Assuming that the wave function varies sufficiently slowly and expanding the 
wave function $\Psi(\mu\pm4\mu_0)$ around $\mu$, we obtain
	\begin{eqnarray}
	 \Psi(\mu\pm4\mu_0) = \Psi(\mu) \pm \frac{\mathrm{d}\Psi(\mu)}{\mathrm{d}\mu}(4\mu_0)
	 +\frac{1}{2}\frac{\mathrm{d}^2\Psi(\mu)}{\mathrm{d}\mu^2}(16\mu_0^2) 
	 \pm \frac{1}{6}\frac{\mathrm{d}^3\Psi(\mu)}{\mathrm{d}\mu^3}(64\mu_0^3) 
	 + O\left(\mu_0^4\frac{\mathrm{d}^4\Psi(\mu)}{\mathrm{d}\mu^4}\right).  \label{wave function expand}
	\end{eqnarray}
Substituting Eq.~(\ref{wave function expand}) into Eq.~(\ref{diff eq_Bojo+CC+V-V expand}), we obtain 
	\begin{subequations}
	\begin{eqnarray}
	\langle\mu|\widehat{C}_{\mathrm{grav}}|\Psi\rangle = 
	 \frac{\sqrt{3}}{\sqrt{\pi^3\gamma^3}l_{\mathrm{Pl}}} \left[ 
	 \frac{\mathrm{d}^2}{\mathrm{d}\mu^2} \bigl(\sqrt{\mu}\Psi(\mu)\bigr) + \frac{\pi\gamma^3l^2_{\mathrm{Pl}}}{9}\mu^{3/2}\Lambda\Psi(\mu) + O(\mu_0)
	 \right] .  \label{continuum limit of diff eq_Bojo+CC}
	\end{eqnarray}
Because $\mu \propto p$, this equation is just the WDW equation as in Appendix~\ref{wdw eq}.
Similarly, we consider the large volume limit of $\langle\mu|\widehat{C}_{\mathrm{grav}}|\Psi\rangle$ 
in the operator ordering (\ref{Hamiltonian_operator_pre2}), 
and then we obtain the WDW equation
	\begin{eqnarray}
	\langle\mu|\widehat{C}_{\mathrm{grav}}|\Psi\rangle = 
	 \frac{\sqrt{3}}{\sqrt{\pi^3\gamma^3}l_{\mathrm{Pl}}} \left[ 
	\sqrt{\mu}\frac{\mathrm{d}^2}{\mathrm{d}\mu^2} \Psi(\mu) + \frac{\pi\gamma^3l^2_{\mathrm{Pl}}}{9}\mu^{3/2}\Lambda\Psi(\mu) + O(\mu_0)
	\right] .  \label{continuum limit of diff eq_Bojo+CC2}
	\end{eqnarray}
	\end{subequations}

Second, we calculate the large volume limit of $\langle v|\widehat{C}_{\mathrm{grav}}|\Psi\rangle$ 
in the operator orderings (\ref{type a}) - (\ref{type d}) in the equi-volume discretisation. 
To do this, we here assume that $|v| \gg 1$ and $\Phi(v)$ varies
sufficiently slowly.
Then, we obtain the WDW equations from the operator orderings (\ref{type a}) - (\ref{type d}), respectively, as follows: 
	\begin{subequations}
	\begin{eqnarray}
&&           \frac{27}{8l_{\mathrm{Pl}}\gamma^{3/2}}\sqrt{\frac{8}{6\pi}} K \Biggl[ \frac{\mathrm{d}^2}{\mathrm{d} v^2} \left(|v| \Phi(v)\right)
 	   + \frac{4\pi}{81}\gamma^3\frac{l^2_{\mathrm{Pl}}}{K^2}\Lambda |v|\Phi(v) + O\left(\frac{\mathrm{d}^3\Phi(v)}{\mathrm{d}v^3}\right)  \Biggr] ,\\
&&           \frac{27}{8l_{\mathrm{Pl}}\gamma^{3/2}}\sqrt{\frac{8}{6\pi}} K \Biggl[ |v| \frac{\mathrm{d}^2}{\mathrm{d} v^2} \Phi(v)
	   + \frac{4\pi}{81}\gamma^3\frac{l^2_{\mathrm{Pl}}}{K^2}\Lambda |v|\Phi(v) + O\left(\frac{\mathrm{d}^3\Phi(v)}{\mathrm{d}v^3}\right)  \Biggr] ,\\
&&           \frac{27}{8l_{\mathrm{Pl}}\gamma^{3/2}}\sqrt{\frac{8}{6\pi}} K \Biggl[ |v| \frac{\mathrm{d}^2}{\mathrm{d} v^2} \Phi(v)
	   + \frac{4\pi}{81}\gamma^3\frac{l^2_{\mathrm{Pl}}}{K^2}\Lambda |v|\Phi(v) + O\left(\frac{\mathrm{d}^3\Phi(v)}{\mathrm{d}v^3}\right)  \Biggr]  ,\\
&&           \frac{27}{8l_{\mathrm{Pl}}\gamma^{3/2}}\sqrt{\frac{8}{6\pi}} K \Biggl[ \frac{\mathrm{d}^2}{\mathrm{d} v^2} \left(|v|\Phi(v)\right) 
	   + \frac{4\pi}{81}\gamma^3\frac{l^2_{\mathrm{Pl}}}{K^2}\Lambda |v|\Phi(v) + O\left(\frac{\mathrm{d}^3\Phi(v)}{\mathrm{d}v^3}\right) \Biggr]  . 
          \end{eqnarray}
	\end{subequations}
\section{Matter Hamiltonian constraint}
\label{sec:matter_hamiltonian}

Here, we shall discuss the matter Hamiltonian operator. 
We assume that the matter Hamiltonian 
constraint can be written by $a^r\epsilon(a, \phi)$ for
arbitrary matter fields~$\phi$,
where $r$ is a constant and $\epsilon(a,
\phi)$ is a function of the matter fields and the scale factor
such that $\epsilon(a, \phi)$ has a nonzero and finite limit
for $a\rightarrow 0$.

First, we consider the case for $r<0$. 
Classically, the matter Hamiltonian diverges for 
$a\to 0$ due to the inverse scale factor.
In LQG, such a divergence can be regularised by Thiemann's
prescription~\cite{Thiemann1}:
we multiply the matter Hamiltonian constraint 
by $1^m=\left(\det e_a^i/\sqrt{\vert\det E\vert}\right)^m$,
where $m$ can be chosen such that we obtain the positive power of the volume factor.
Similarly, we regularise the inverse scale factor in LQC as follows~\cite{vand}. 
The classical Poisson bracket takes the form 
	\begin{equation}
	 \left\{c, V^{\frac{2l}{3}}\right\}=\mathrm{sgn}(p)\frac{8\pi\gamma Gl}{3}  \vert p\vert^{l-1},
	\end{equation}
where we have used $V=|p|^{3/2}$, and $l$ is the ambiguity parameter similar to $m$ in LQG.
Notice that if  we choose $0<l<1$, the right hand side denotes the inverse power of $p$, 
while the left hand side involves the positive power of the volume. 
Using this property, we rewrite the inverse scale factor $a^{-1}$ as 
	\begin{equation}
	 a^{-1}=\frac{V_0^{\frac{1}{3}}}{\sqrt{\vert p\vert}} 
         =\left(\frac{3\mathrm{sgn}(p)}{8\pi\gamma Gl}\left\{c, V^{\frac{2l}{3}}\right\}\right)^{\frac{1}{2(1-l)}}V_0^{\frac{1}{3}}, 
	\end{equation}
where we used $\vert p\vert =V_0^{2/3}a^2$. 
Then, the matter Hamiltonian becomes
	\begin{eqnarray}
         \nonumber C_{\mathrm{matter}}&=&a^{r}\epsilon(a,\phi)
         =\left[\frac{1}{\sqrt{\vert p\vert} }\right]^{-r}V_0^{-\frac{r}{3}}\epsilon(a, \phi)\\
         &=&\left[\frac{3\mathrm{sgn}(p)}{8\pi\gamma Gl}\left\{c, V^{\frac{2l}{3}}\right\}\right]^{-\frac{r}{2(1-l)}}V_0^{-\frac{r}{3}}\epsilon(a, \phi). 
	\end{eqnarray}
This classical formula can be represented exactly in terms of holonomies as follows: 
	\begin{equation}
         C_{\mathrm{matter}}=
         \left[\frac{\mathrm{sgn}(p)}{4\pi\gamma Gl\bar{\mu}}
         \mathrm{Tr}\left(\sum_i\tau^ih_i^{(\bar{\mu})}\left\{h_i^{(\bar{\mu})}{}^{-1}, V^{\frac{2l}{3}}\right\}\right)\right]
         ^{-\frac{r}{2(1-l)}}V_0^{-\frac{r}{3}}\epsilon(a, \phi), 
	\end{equation}
and we can quantise this Hamiltonian immediately by replacing the Poisson bracket $\{\bullet ,\bullet \}$ with $-i[\bullet ,\bullet ]$. 
Using $G=l_{\mathrm{Pl}}^2$ and Eq.~(\ref{eq:mubar}), after some calculation, 
we obtain 
	\begin{equation}
         \widehat{C}_{\mathrm{matter}}=
         \left[-\frac{3i(\mathrm{sgn}(p))}{4\pi\gamma l_{\mathrm{Pl}}^2l}\left(\frac{6}{8\pi l_{\mathrm{Pl}}^2\gamma}\right)^{1/2}\left(\frac{K\widehat{V}}{\sqrt{3}}\right)^{\frac{1}{3}}
          \widehat{\biggl(\sin\left(\frac{\bar{\mu}c}{2}\right)V^{\frac{2l}{3}}\cos\left(\frac{\bar{\mu}c}{2}\right)
          -\cos\left(\frac{\bar{\mu}c}{2}\right)V^{\frac{2l}{3}}\sin\left(\frac{\bar{\mu}c}{2}\right)\biggr)}\right]^{-\frac{r}{2(1-l)}}
          V_0^{-\frac{r}{3}}\widehat{\epsilon(a, \phi)}. 
	\end{equation}
Then, the matter Hamiltonian operator acts on $\vert v\rangle$ as 
        \begin{equation}\label{matter constraint0}
         \widehat{C}_{\mathrm{matter}}\vert v\rangle
         = 
         \left[\frac{3\mathrm{sgn}(p)}{8\pi\gamma l_{\mathrm{Pl}}^2l}
         \left(\frac{6}{8\pi l_{\mathrm{Pl}}^2\gamma}\right)^{1/2}\left(\frac{KV_v}{\sqrt{3}}\right)^{\frac{1}{3}}
          \left(V_{v+1}^{\frac{2l}{3}}-V_{v-1}^{\frac{2l}{3}}\right)
          \right]^{-\frac{r}{2(1-l)}}V_0^{-\frac{r}{3}}\epsilon_0(a, \phi)\vert v\rangle, 
	\end{equation}
where $V_v=\left(8\pi\gamma l_{\mathrm{Pl}}^2/6\right)^{3/2}\vert v\vert/K$ and
$\epsilon_0(a, \phi)$ is an eigenvalue of $\widehat{\epsilon(a, \phi)}$.
Therefore, $\vert v\rangle$ is an eigenstate of $\widehat{C}_{\mathrm{matter}}$,
and its eigenvalue vanishes for $v=0$.
Therefore, substituting this into Eq.~(\ref{eq:matter}), 
since the eigenvalue of the matter Hamiltonian operator vanishes for $v=0$, 
we can uniquely determine the coefficients $\Phi(v,\phi)$ except for
$\Phi(0,\phi)$.
In conclusion, the matter Hamiltonian operator does not affect
the absence or presence of the singularity for $r<0$.

Second, we consider the case with $r=0$.
In this case, the matter Hamiltonian is given by 
$C_{\mathrm{matter}}=\epsilon(a, \phi)$. 
Thus, the matter Hamiltonian operator acts on the state $\vert v\rangle$ as
\begin{equation}
         \widehat{C}_{\mathrm{matter}}\vert v\rangle = \epsilon_0(a,\phi)\vert v\rangle.
        \end{equation}
Then, $\vert v\rangle$ is an eigenstate of $\widehat{C}_{\mathrm{matter}}$
but in this case its eigenvalue does not vanish even for $v=0$.
Substituting this into Eq.~(\ref{eq:matter}), we find that 
we cannot uniquely determine $\Phi(0,\phi)$ from this equation
and hence cannot obtain the unique wave function $\Phi(v,\phi)$ beyond $v<0$.
Therefore, in this case, even though there is no initial singularity
without matter fields, there appears the initial singularity due to 
the presence of the matter fields.

Finally, for $r>0$, the matter Hamiltonian 
becomes $C_{\mathrm{matter}}=V^{r/3}\epsilon(a, \phi)$. 
Thus, the matter Hamiltonian 
operator acts on the state $\vert v\rangle$
as
\begin{equation}
         \widehat{C}_{\mathrm{matter}}\vert v\rangle =
V_v^{\frac{r}{3}}\epsilon_0(a, \phi)\vert v\rangle.
        \end{equation}
Therefore, $\vert v\rangle$ is an eigenstate of $\widehat{C}_{\mathrm{matter}}$,
and its eigenvalue vanishes for $v=0$.
Similarly to the case of $r<0$, the matter Hamiltonian does
not affect the absence or presence of the singularity.

\end{document}